\documentclass[twocolumn,amsmath,amssymb,pra]{revtex4-1}

\usepackage{graphicx}
\usepackage{color}
\usepackage{dcolumn}
\usepackage{amsmath}
\usepackage{CJK}
\usepackage{subfigure}
\usepackage{sidecap}

\usepackage{mathrsfs}
\usepackage{amsfonts}
\usepackage{indentfirst}
\usepackage{bm}

\newcommand{\be}{\begin{equation}}
\newcommand{\ee}{\end{equation}}
\newcommand{\bey}{\begin{eqnarray}}
\newcommand{\eey}{\end{eqnarray}}
\newcommand{\bw}{\begin{widetext}}
\newcommand{\ew}{\end{widetext}}

\newcommand{\ra}{\rangle}
\newcommand{\la}{\langle}

\newcommand{\ba}{\begin{array}}
\newcommand{\ea}{\end{array}}
\newcommand{\bi}{\begin{itemize}}
\newcommand{\ei}{\end{itemize}}
\newcommand{\bem}{\begin{enumerate}}
\newcommand{\eem}{\end{enumerate}}

\begin{document}

\title{Probing an excited-state quantum phase transition in a quantum many-body system via an out-of-time-order
       correlator}

\author{Qian Wang$^{1,2}$}  

\affiliation{$^{1}$Department of Physics, Zhejiang Normal University, Jinhua 321004, China \\
$^2$Center for Theoretical Physics of Complex Systems, Institute for Basic Science, Daejeon 34051, Korea}

\author{Francisco P\'erez-Bernal$^{3,4}$}

\affiliation{$^{3}$Departamento de Ciencias Integradas y Centro de Estudios Avanzados en F\'{\i}sica, Matem\'aticas y Computaci\'on, Universidad de Huelva, Huelva 21071, Spain\\$^{4}$Instituto Carlos I de F\'{\i}sica Te\'orica y Computacional, Universidad de Granada, Granada 18071, Spain}

\date{\today}

\begin{abstract}
  Out-of-time-order correlators (OTOCs) play an increasingly important
  role in different fields of physics and in particular they provide
  a way of quantifying information scrambling in quantum many-body
  systems. We verify that an OTOC can be used to probe an
  excited-state quantum phase transition (ESQPT) in a quantum
  many-body system.  We examine the dynamical properties of an OTOC in
  the Lipkin-Meshkov-Glick model, which undergoes an ESQPT,
  using the exact diagonalization method.  We show that the long-time
  evolution of the proposed OTOC is remarkably different in the
  different phases of the ESQPT. In consequence, we put the
  long-time averaged value of the OTOC forward as a possible ESQPT order
  parameter.  Our results highlight the connections between OTOCs and
  ESQPTs, opening the possibility of using OTOCs for accessing
  experimentally ESQPTs in quantum many-body systems.
\end{abstract}

\maketitle

\section{Introduction}

The concept of an excited-state quantum phase
transition (ESQPT) was introduced as a generalization of the 
ground-state quantum phase transition (QPT) 
\cite{Carr2010} to
the realm of excited states. Their occurrence is associated with a
singular behavior of the excited-state level density at a given critical energy, once a
Hamiltonian control parameter passes through a critical value associated with the ground-state QPT \cite{Cejnar2006,Caprio2008}. This
divergence of the density of states at the critical energy in the
mean-field limit is the most remarkable feature of ESQPTs
\cite{Cejnar2006,Caprio2008,Brandes2013,Stransky2014,Santos2016,Santos2015,PBernal2017}.
In recent years, ESQPTs have been studied in several quantum
many-body models (see, e.g., Ref.~\cite{Santos2016} and references
therein) and have also been identified experimentally in different
systems \cite{Larese2011, Larese2013,Dietz2013,Zhao2014, KRivera2019}.
Furthermore, the impact of an ESQPT in the system
nonequilibrium dynamics has attracted much attention
\cite{Puebla2013,Puebla2015,Engelhardt2015,Kopylov2015,Kloc2018}.  It
has been found that the evolution of isolated systems
can be substantially slowed down \cite{Santos2016,Wang2019},
decoherence processes in open systems can be enhanced
\cite{Relano2008,PFernandez2009}, and changes in the quantum work
distribution can occur \cite{Wang2017}  under an ESQPT.  In particular, an abrupt
increase of the entropy at the critical point of the ESQPT has been
revealed \cite{Lobez2016}, which means that ESQPTs strongly
influence the propagation of quantum information in many-body
systems.

The out-of-time-order correlator (OTOC) was originally
introduced by Larkin and Ovchinnikov in the context of superconductivity studies
\cite{Larkin1969}. It has recently  been rekindled \cite{kitaev}
because it offers an interesting perspective on the occurrence of
quantum chaos and on the exploration of information propagation in
quantum many-body systems (see, e.g.,~\cite{Swingle2018,Roberts2016} and references therein). Possible
OTOC realizations are currently being investigated in many fields, offering different and interesting insights into physical systems. Some of the most important results
involve the dynamics of quantum information
\cite{Roberts2016,Luitz2017,Iyoda2018,Keyserlingk2018,Niknam2018,LSwan2019,Yahya2019}.
The OTOC study has renewed the interest in the correspondence between
classical and quantum chaos
\cite{Maldacena2016,Hosur2016,Kukuljan2017,Hashimoto2017,Rozenbaum2017,Rozenbaum2019,GMata2018,Jalabert2018,Fortes2019,THerrera2018,CCarlos2019,Cameo2019}
with some analytical breakthroughs in the field of high-energy physics, mostly regarding the black hole information problem
\cite{Lashkari2013,Shenker2014, Maldacena2017} and the
Sachdev-Ye-Kitaev model \cite{Polchinski2016, Maldacena2016b}. The OTOCs
have also found application in condensed-matter systems (see, e.g., 
Refs.~\cite{Swingle2017,Patel2017,Patel2017b,Dora2017,Shen2017,Sunz2018,Heyl2018,Huang2017,Fan2017,Sahu2018})
as well as in statistical physics
\cite{Campisi2017,Chenu2018,Chenu2019}.
It has been verified that OTOCs can be employed
for the characterization of phase transitions in quantum many-body systems,
e.g.,~ground-state QPTs \cite{Shen2017,Heyl2018}, many-body localization
transitions \cite{Huang2017}, ergodic-nonergodic
transitions \cite{Buijsman2017,Ray2018}, and dynamical phase transitions
\cite{Heyl2018}. 
Moreover, recent progress in the experimental detection of quantum
correlations and in quantum control techniques applied to systems as atoms,
molecules, or photons has led to the direct observation of an OTOC in
spin  \cite{Li2017,Wei2018} and trapped ions
\cite{Garttner2017} systems.
Finally, a relationship between the OTOC for
projection operators and the number of principal components
(participation ratio) has been found recently
\cite{Borgonovi2019}. The latter is another quantity that has proved very
relevant in the study of ESQPTs in two-level systems \cite{Santos2015}. 
In addition, the fact that OTOCs cast light upon ESQPTs can be expected 
if one considers that there exists a clear link between OTOCs and Loschmidt echoes
\cite{Peres1984}, something that is clear  from the OTOC
definition provided below, and Loschmidt echoes are known to be strongly influenced by ESQPTs
\cite{PFernandez2009,Wang2017}.

Therefore, we zero in on revealing how the signatures of an ESQPT modify the dynamical properties of OTOCs, which are defined as follows \cite{Hashimoto2017,Swingle2018}.  Given a system with a Hamiltonian $H$, an initial state, and two operators ${W}$ and ${V}$, the spread with time of the operator ${W}$ can be probed through the expectation value of the squared module of a commutator with a second operator ${V}$,
\begin{align} \label{OTOC_0}
C_{w,v}(t)&=\la[{W}(t),{V}(0)]^\dag[{W}(t),{V}(0)]\ra  \notag \\
  &=2\operatorname{Re}[A_{w,v}(t)]-2\operatorname{Re}[F_{w,v}(t)]~,
\end{align}
where ${W}(t)$ is the operator ${W}$ in the Heisenberg representation
${W}(t) = e^{i H t} W e^{-i H t}$. The commutator module is split into two terms. The first term
$A_{w,v}(t)=\la V^\dag(0)W^\dag(t)W(t)V(0)\ra$ has the same time order
as the usual response functions and it is a quantity that tends to a constant
value for long times \cite{Heyl2018,Gu2016}. The second term
\be \label{OTOC_1}
F_{w,v}(t)=\la{W}^\dag(t){V}^\dag(0){W}(t){V}(0)\ra~,  \ee
\noindent has a particular
out-of-time order, with a nontrivial time dependence. Therefore, the
long-time dynamical properties of $C_{w,v}(t)$ are only determined
by $F_{w,v}(t)$. Due to the special
time ordering implicit  in Eq.~(\ref{OTOC_1}), $F_{w,v}(t)$ is 
dubbed an OTOC.  If  $W$ and $V$ are unitary operators,
$C_{w,v}(t)$ in Eq.~(\ref{OTOC_0}) can be further simplified to
$C_{w,v}(t)= 2-2\operatorname{Re}[F_{w,v}(t)]$. In this case,
the $C_{w,v}(t)$ time dependence is fully dependent on $F_{w,v}(t)$.  In our case, the operators initially commute,
$[{W}(0), {V}(0)]=0$, but this is not a crucial restriction. In the literature, the
average $\la\cdot\ra$in the above-mentioned quantities was performed over the canonical ensemble, but, as in other recent
works, we perform an average over initial states.

In this work we investigate the time evolution after a sudden quench
of an OTOC for an isolated quantum many-body system which undergoes an
ESQPT, the Lipkin-Meshkov-Glick (LMG) model \cite{Lipkin1965}.  Our
results indicate that the selected OTOC can be considered as a
candidate for the order parameter of the underlying ESQPT due to the
drastic changes that occur when the quench drives the system across
the ESQPT critical point.  We confirm this result by examining the
microcanonical OTOC, studying the time dependence for initial
eigenstates above and below the critical energy of the ESQPT, and
confirming that the OTOC evolution exhibits a sudden change when
crossing the critical energy, making it possible to classify the ESQPT
phases according to the OTOC long-time averaged value.


 \begin{figure}
  \includegraphics[width=\columnwidth]{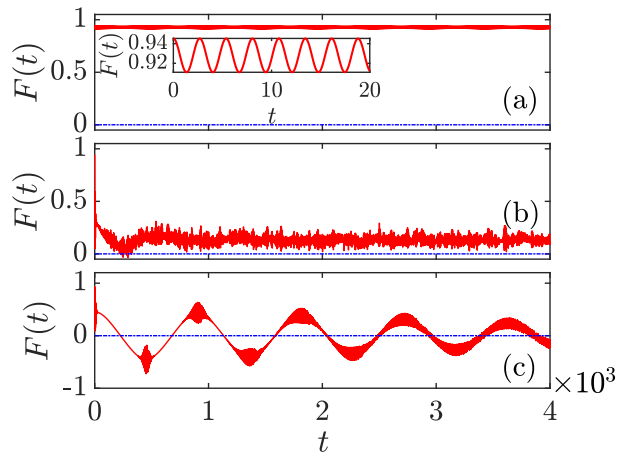}
  \caption{(Color online) Time evolution of the OTOC $F(t)$ for three different $\lambda$ values in the LMG model: 
  (a) $\lambda=0.1$, (b) $\lambda=1$, and (c) $\lambda=2$. Red solid (blue dot-dashed) lines are the
   real (imaginary) part of OTOC.  The control parameter value is $\alpha=0.4$ and the system size $N=400$.
   The inset in (a) shows a close-up in the evolution of $F(t)$ within $0\leq t\leq 20$.
   All quantities are unitless.}
  \label{Tem}
 \end{figure}

 \section{The Lipkin-Meshkov-Glick model} \label{MLMG}
 
 The LMG model describes an 
 Ising spin chain with infinite-range interactions.
 The collective spin operators are defined as $S_{\gamma}=\sum_{i=1}^N\sigma^i_\gamma$, with $\gamma=x,y,z$,
 where $N$ is the total number of spins and $\sigma_\gamma^i$ are the Pauli spin matrices for the $i$th spin. 
 The Hamiltonian of the LMG model can be written as (see, e.g.,~Ref.~\cite{Santos2016})

 \be \label{LMG}
  H=-\frac{2(1-\alpha)}{\mathcal{S}}S_x^2+\alpha(S_z+\mathcal{S})~,
 \ee
 
 \noindent where the
 total spin of the system is a conserved quantity, i.e.,
 $[\mathbf{S}^2,H]=0$. Therefore, we may restrict our study to the
 sector $\mathcal{S}=N/2$, reducing the Hilbert space to a 
 dimension $\mathcal{D}_H=N+1$. The control parameter is $\alpha$, which takes values in the range $0\le \alpha\le 1$.

 It is known that, in the mean-field (or thermodynamic) limit, i.e.,
 when $N\to\infty$, the LMG model exhibits a second-order QPT at a
 critical value of the control parameter $\alpha_c=0.8$
 \cite{Dusuel2004,Leyvraz2005,ocasta1}.  It is worth pointing out that
 recent studies have shown that an OTOC can be a good probe for the
 ground-state QPT in the Lipkin model \cite{Heyl2018}.  The $\la S_x\ra$ 
 parameter acts as an order parameter for this ground-state phase
 transition. The phase with $\alpha<\alpha_c$ is the broken-symmetry
 phase, where $\la S_x\ra\propto|\alpha-\alpha_c|^{1/2}$, while for $\alpha>\alpha_c$, the phase is the
 symmetric phase with $\la S_x\ra=0$ \cite{Botet1983}.  However, besides the ground-state QPT,
 the LMG model also undergoes an ESQPT with a critical energy $E_c=0$
 \cite{Relano2008,PFernandez2009}.  In the following we
 focus on the identification of the signatures of this ESQPT making use of
 an OTOC.

 \begin{figure*}
   \begin{minipage}{0.46\linewidth}
   \centering
   \includegraphics[width=\columnwidth]{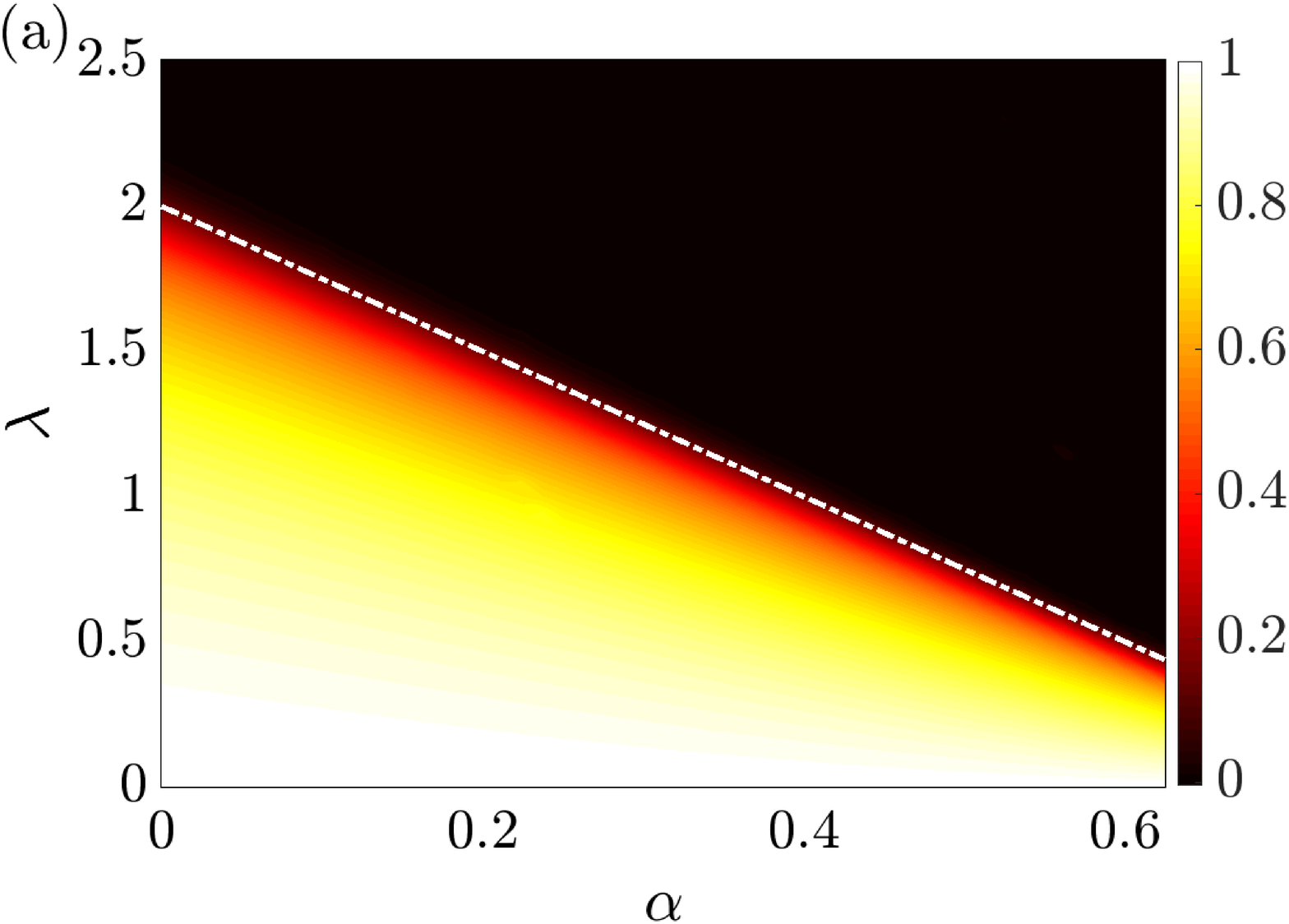}
   \end{minipage}
   \begin{minipage}{0.46\linewidth}
   \centering
   \includegraphics[width=\columnwidth]{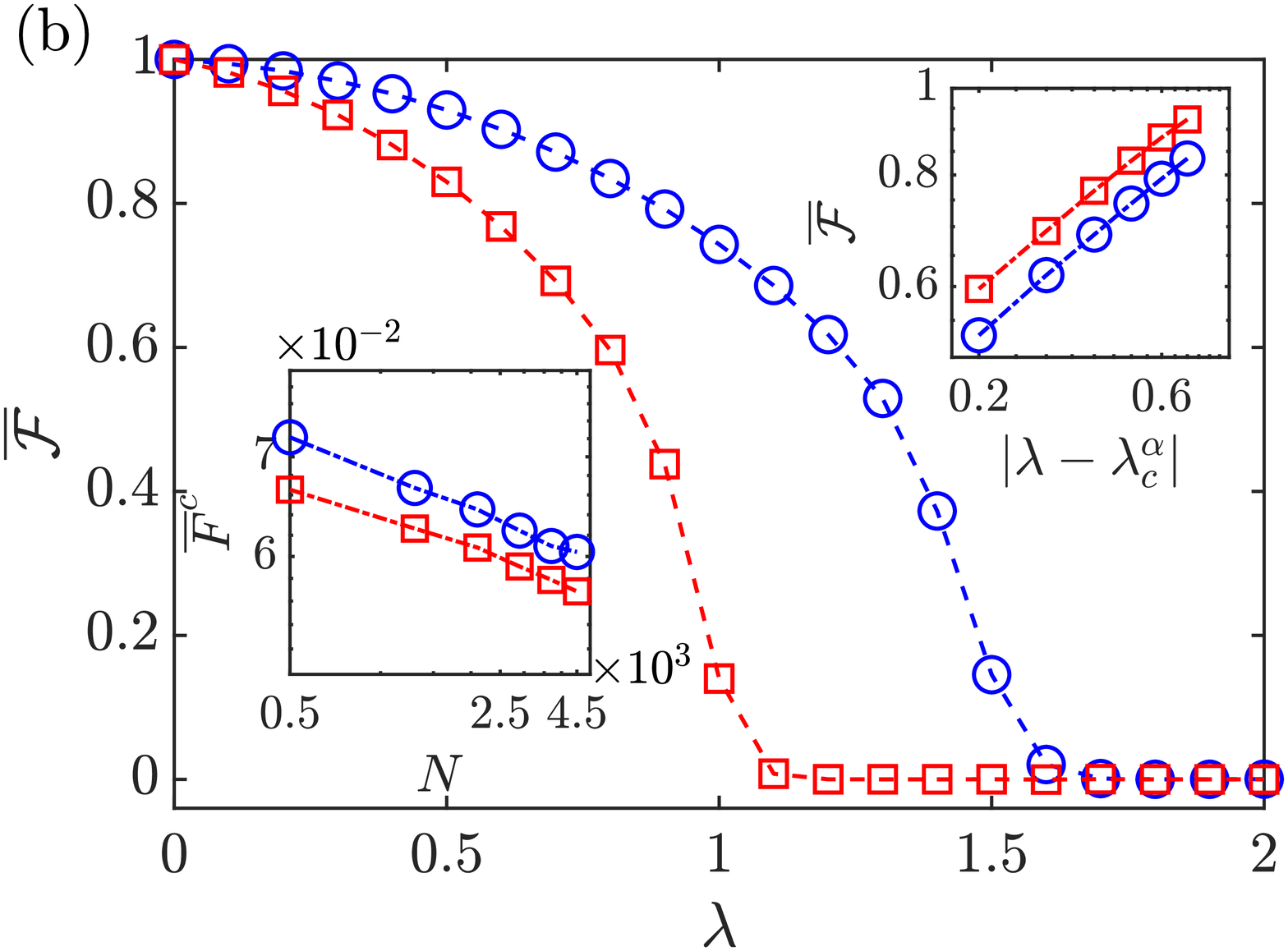}
   \end{minipage}
   \caption{(Color online) (a) Heat map depicting the normalized
     long-time averaged OTOC value $\overline{\mathcal{F}}$ as a
     function of $\alpha$ and $\lambda$ for $N=400$.  The white
     dot-dashed line indicates the critical value of $\lambda$,
     obtained from Eq.~(\ref{criticalV}).  (b) Normalized long-time
     averaged OTOC value $\overline{\mathcal{F}}$ as a function of
     $\lambda$ for $\alpha = 0.2$ (blue circles) and $0.4$ (red
     squares) for $N=400$.  The left inset shows the critical value
     $\overline{F}^c=\overline{F}(\lambda_c)$ as a function of the
     system size $N$ (on a log-log scale) for
     $(\alpha,\lambda_c) = (0.4, 1.0)$ (red squares) and
     $(\alpha,\lambda_c) = (0.2, 1.5)$ (blue circles).  The right inset shows
     $\overline{\mathcal{F}}$ as a function of
     $|\lambda-\lambda_c^\alpha|$ (on a log-log scale) in the
     neighborhood of the critical point for a system size $N=400$ and
     control parameter values $\alpha=0.2$ (blue circles) and
     $\alpha=0.4$ (red squares).  In both panels
     $\overline{\mathcal{F}}$ is obtained normalizing $\overline{F}$,
     for each $\alpha$ value, by its $\lambda = 0$ initial value,
     i.e.,~$\overline{\mathcal{F}}=\overline{F}/\overline{F}(\lambda=0)$.
     All quantities are unitless.}
   \label{otcphase}
 \end{figure*}

 In our study we follow the  approach of Ref.~\cite{Heyl2018}, using the rescaled order parameter of the QPT in the LMG model as the OTOC operators (\ref{OTOC_1}), i.e., ${V}={W}=S_x/\mathcal{S}$.
Initially the system is prepared in the ground state of the Hamiltonian (\ref{LMG}) and we consider the following sudden quench process. At $t=0$, an external magnetic field along the $z$ direction is added, with strength $\lambda$; then we investigate numerically the evolution of the OTOC defined in Eq.~(\ref{OTOC_1})
 under the post quench Hamiltonian $H_f=H+\lambda S_z$, where $H$ is provided by Eq.~(\ref{LMG}).
 The evolution of the operator $W$ in the Heisenberg representation can be expressed as $W(t)=e^{iH_ft}We^{-iH_ft}$.

 Varying the strength of the external field, one can drive the system through the
 critical energy of the ESQPT.  The critical strength $\lambda_c$, which takes the LMG system to the critical energy $E_c=0$,
 can be easily obtained through the semiclassical approach. The final result reads \cite{Relano2008}
 \be \label{criticalV}
   \lambda_c=\frac{1}{2}(4-5\alpha)~,
 \ee
 where $\alpha$ is such that $\alpha<\alpha_c=0.8$.

 To calculate numerically the OTOC in the LMG model, we diagonalize
 the pre- and post quench Hamiltonians of the model in the basis
 $|\mathcal{S},m_x\ra$, where $-\mathcal{S}\leq m_x\leq \mathcal{S}$.  In this basis,
 the non-zero elements of the Hamiltonian  matrix (\ref{LMG}) are
 \be
  \begin{gathered}
   \la\mathcal{S},m_x|H|\mathcal{S},m_x\ra=-\frac{2(1-\alpha)}{\mathcal{S}}m_x^2+\alpha\mathcal{S}, \notag \\
   \la\mathcal{S},m_x-1|H|\mathcal{S},m_x\ra=\frac{\alpha}{2}\sqrt{\mathcal{S}(\mathcal{S}+1)-m_x(m_x-1)}. \notag
  \end{gathered}
 \ee

 \section{Results} \label{NRS}
 
 We present numerical simulation results for the OTOC (\ref{OTOC_1})
 in the LMG model obtained by the exact diagonalization of the
 Hamiltonian.  In Fig.~\ref{Tem} we plot the OTOC time evolution for
 a system size $N = 400$, $\alpha=0.4$, and different values of
 $\lambda$.  According to Eq.~(\ref{criticalV}), the critical value of
 the external field strength in this case is $\lambda_c=1$.  Several
 remarkable features can be observed in Fig.~\ref{Tem}.  First, the
 imaginary part of $F(t)$ is always zero, regardless of the $\lambda$
 values.  Second, the behavior of the real part of the OTOC, which we
 denote by $F_R(t)$, clearly depends on whether the value of $\lambda$
 is below or above the critical value $\lambda_c=1$.  Specifically, as
 shown in Fig.~\ref{Tem}(a), $F_R(t)$ is characterized by a 
 small-amplitude oscillation around a large positive value for
 $\lambda<\lambda_c$ [see the inset in
 Fig.~\ref{Tem}(a)]. Figure.~\ref{Tem}(b) displays the system evolution
 for $\lambda=\lambda_c=1$, where $F_R(t)$, after a fast decrease to
 its minimum value, irregularly oscillates around a small nonzero
 positive value.  However, when $\lambda>\lambda_c$, the oscillation
 pattern radically changes, transforming into a damped oscillation, as
 can be seen in Fig.~\ref{Tem}(c).  Once enough time passes, $F_R(t)$
 reaches zero and the time-independent steady-state value of $F(t)$
 equals zero.

 \begin{figure}
  \includegraphics[width=\columnwidth]{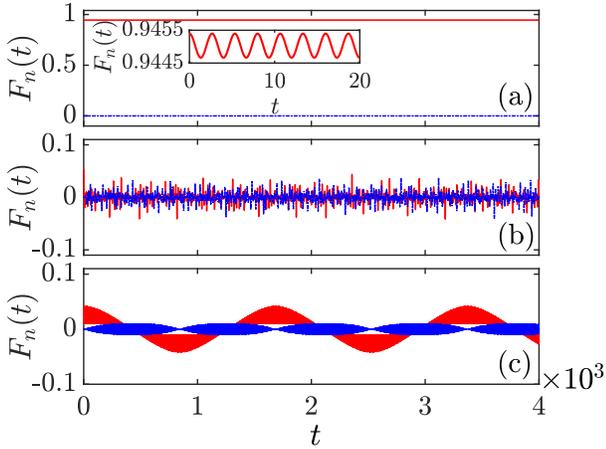}
  \caption{Time evolution of the LMG microcanonical OTOC $F_n(t)$ for different
   initial states: (a) $|n=0\ra$ with an energy $E_n/N=-0.4167$,
   (b) $|n=149\ra$ with $E_n/N=-6.5419\times10^{-4}$, and (c) $|n=219\ra$
   with $E_n/N=0.1467$. Red solid lines are the real part of $F_n(t)$ and blue dot-dashed
   lines denote the imaginary part of $F_n(t)$.
   Other parameter values are $\alpha=0.4$ and $N=300$.
   The inset in (a) shows a close-up of the evolution of $F_n(t)$ with $0\leq t\leq20$.
   The axes in all figures are unitless.}
  \label{Etmc}
 \end{figure}

 The above-mentioned results confirm our assumption that the existence of an ESQPT 
 in the system has a strong impact on the OTOC  dynamical properties.
 In fact, the ESQPT is clearly disclosed by the singular
 behavior of the OTOC at the critical strength of the external field $\lambda_c$.
 In addition, states with energies above or under the critical energy value, in the different phases of the ESQPT, 
 can be distinguished by the remarkable distinct OTOC dynamical behavior.
 Moreover, Fig.~\ref{Tem} evidences that the steady-state value of $F(t)$
 is finite when $\lambda<\lambda_c$, while it is zero for $\lambda>\lambda_c$.
 To clarify this, we study the steady-state value of $F(t)$, which can be
 obtained by calculating the long-time averaged value of $F_R(t)$,
 \be \label{AvgF}
    \overline{F}=\lim_{T\to\infty}\frac{1}{T}\int_0^T F_R(t)dt~,
 \ee
 where $T$ is the total evolution time.

 The aforementioned features of $F(t)$ strongly indicate that $\overline{F}$ can be identified
 as an order parameter for the ESQPT in the LMG model.
 To verify this statement, we evaluate $\overline{F}$ in the LMG model, obtaining the results shown in Fig.~\ref{otcphase}.
 It is worth mentioning that, in order to capture all the intricacies of $F_R(t)$ in the numerical simulation,
 the total evolution time $T$ in Eq.~(\ref{AvgF}) should be large enough.
 In the present work, the total evolution time is $T=10^4$, and we have checked that  the results obtained 
 for larger $T$ values are qualitatively similar.  We define the quantity 
 $\overline{\mathcal{F}}=\overline{F}/\overline{F}(\lambda=0)$, which is obtained by 
 normalizing the long-time averaged value  of $F_R(t)$ by the $\lambda = 0$ value of $\overline{F}$. 
 This is the quantity depicted in Figs.~\ref{otcphase}(a) and \ref{otcphase}(b).


 
 \begin{figure}
   \begin{minipage}{\linewidth}
   \centering
   \includegraphics[width=\columnwidth]{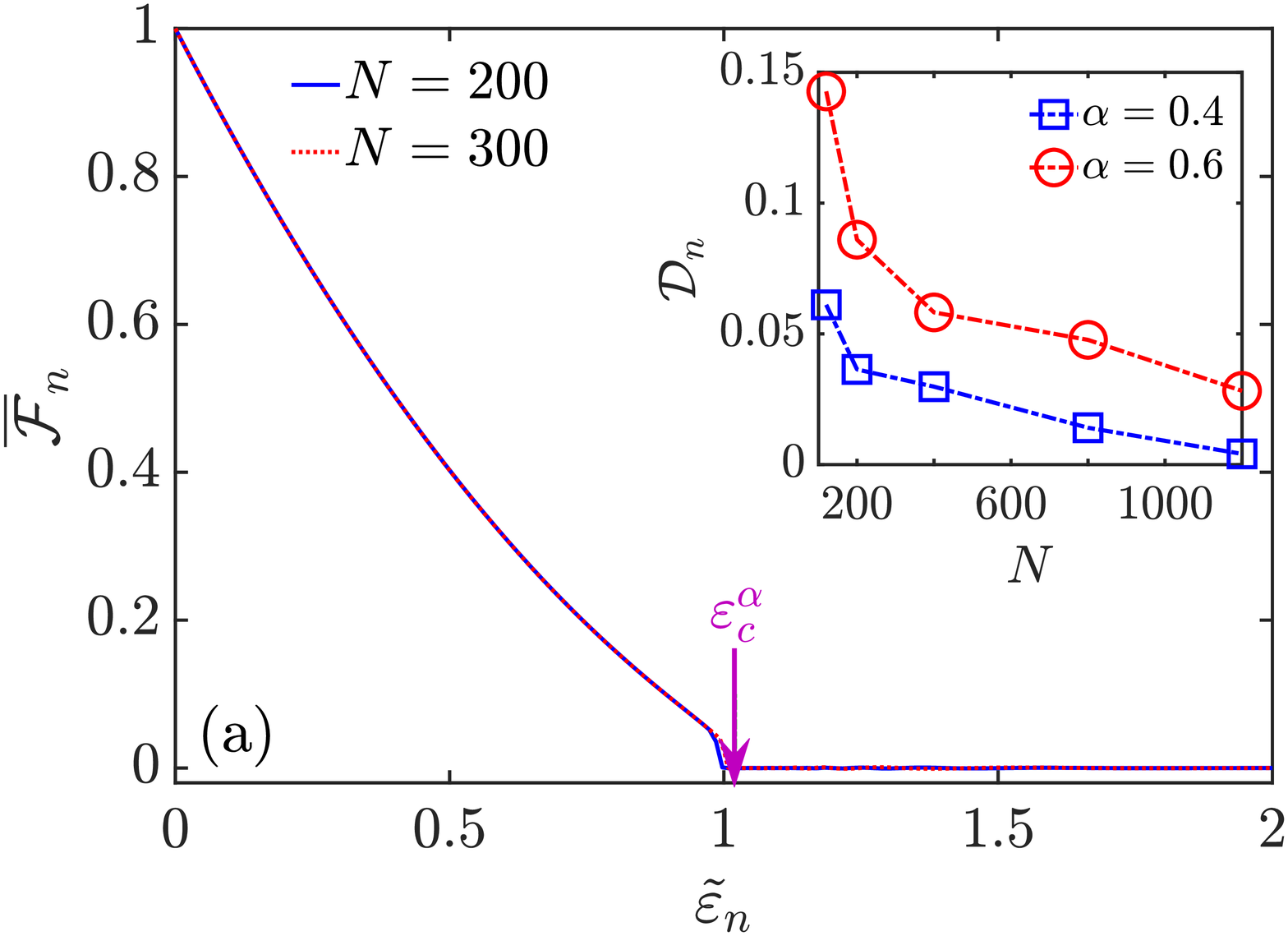}
   \end{minipage}
   \begin{minipage}{\linewidth}
   \centering
   \includegraphics[width=\columnwidth]{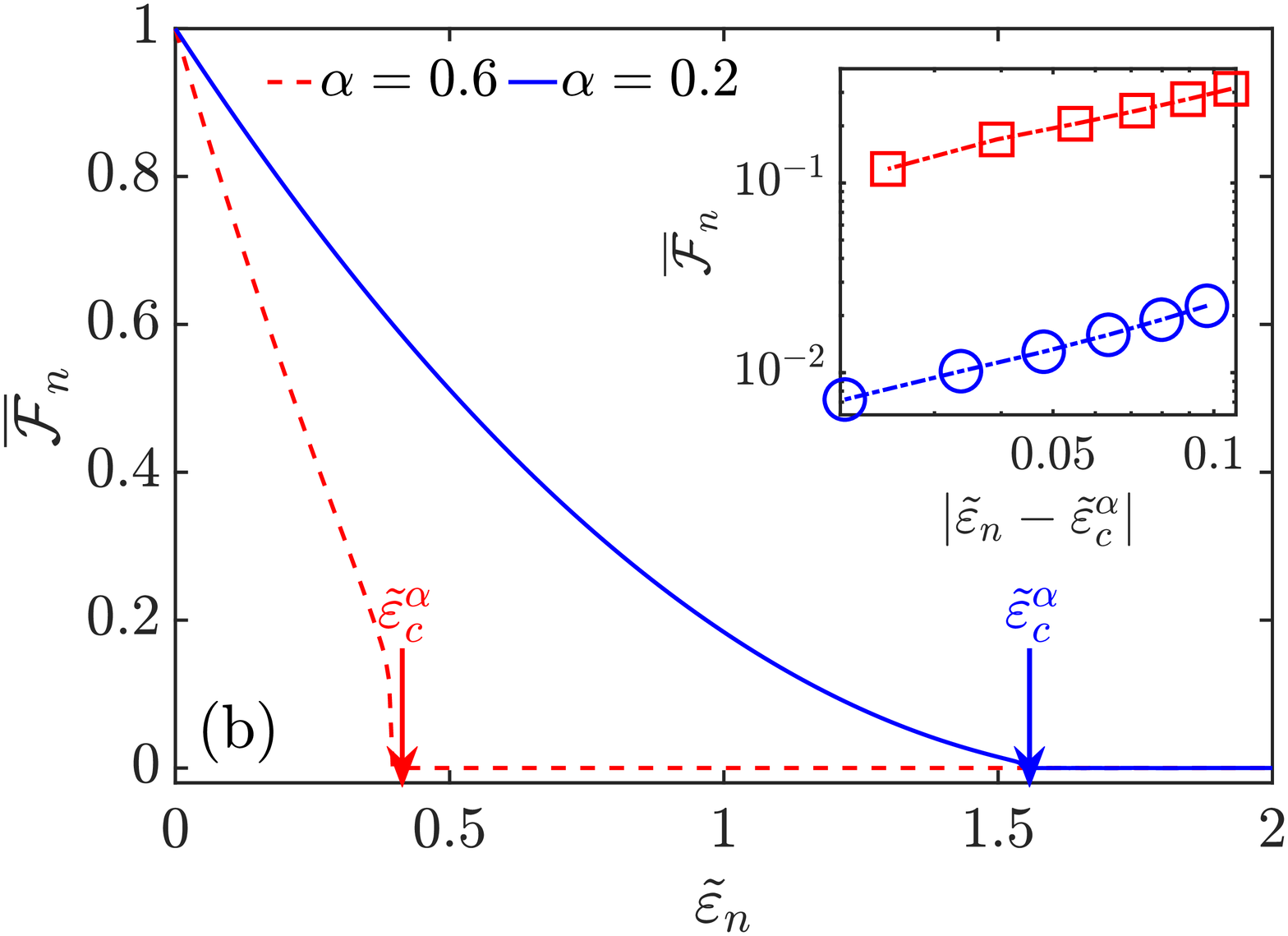}
   \end{minipage}
   \caption{(a) Normalized long-time averaged value of the microcanonical OTOC $\overline{\mathcal{F}}_n$ as a
     function of the rescaled energy $\tilde{\varepsilon}_n$ for different system
     size values $N$ with $\alpha=0.4$. The purple arrow indicates the rescaled 
     critical energy of the ESQPT with $\alpha=0.4$.
     The inset shows the variation of $\overline{\mathcal{F}}_n$'s increase or decrease amplitude $\mathcal{D}_n$ 
     (see the text for its definition) near the critical energy with the system size $N$. 
     (b) Normalized long-time averaged value of the microcanonical OTOC $\overline{\mathcal{F}}_n$
     as a function of the rescaled energy $\tilde{\varepsilon}_n$ for
     different values of $\alpha$ with $N=300$. The blue and red arrows indicate the rescaled critical energy
     $\tilde{\varepsilon}_c^\alpha$ for $\alpha=0.2$ and $\alpha=0.6$, respectively. 
     The inset shows $\overline{\mathcal{F}}_n$ as a function of
     $|\tilde{\varepsilon}_n-\tilde{\varepsilon}^\alpha_c|$ (on a log-log scale) in the neighborhood of the critical energy for
     $\alpha=0.2$ (blue circles) and $\alpha=0.6$ (red squares).
     In both panels the long-time averaged value of the microcanonical OTOC has been normalized by its initial value for
     the ground state $n=0$, i.e., $\overline{\mathcal{F}}_n=\overline{F}_n/\overline{F}_0$, 
     while the eigenenergies are rescaled as
     $\tilde{\varepsilon}_n=2(E_n-E_0)/(E_{max}-E_0)$, where $E_0$ is the ground-state energy and 
     $E_{max}$ is the highest eigenstate energy.
     All quantities are unitless.}
   \label{otceng}
 \end{figure}


 In Fig.~\ref{otcphase}(a) we present the normalized long-time averaged value of the OTOC $\overline{\mathcal{F}}$ as a function of the  $\alpha$ and $\lambda$ parameters for a system size $N=400$.
 We first note that $\overline{\mathcal{F}}$ is a continuous function of $\alpha$ and $\lambda$ and it drops to zero once the white dot-dashed line that indicates the  critical values $\lambda_c$ is crossed. In Fig.~\ref{otcphase}(b) we depict vertical cuts of Fig.~\ref{otcphase}(a) for $\alpha =0.2$ and $0.4$. In both cases, $\overline{\mathcal{F}}$ decreases and vanishes once $\lambda$ reaches the critical value $\lambda_c$. This is the expected dependence for a ground-state QPT order parameter and therefore an analysis of the scaling properties of the long-time averaged OTOC value is germane to this discussion.

 From the behavior of $\overline{\mathcal{F}}$ in both panels of Fig.~\ref{otcphase}, 
 it seems reasonable to expect that the value of
 $\overline{F}$ when $\lambda = \lambda_c$, denoted by $\overline{F}^c$, 
 scales with the system size in such a way that it tends to zero in the thermodynamic limit. 
 In order to check this point, we have included the left inset of Fig.~\ref{otcphase}(b), 
 where $\overline{F}^c$  is depicted as a function of the system size 
 $N$ for $\lambda_c = 1.0$ and $1.5$ on a log-log scale. Using a least-squares fit, 
 we find that the $N$ dependence of $\overline{F}^c$ follows a power law 
 $\overline{F}^c\propto N^{-\mu}$ for both $\lambda_c$ values, with a finite-size scaling exponent $\mu\approx0.084(8)$.
 A second aspect of interest is what happens, for a given system size, when 
 $\lambda\to\lambda_c^\alpha$. We have depicted in the right inset of Fig.~\ref{otcphase}(b) the dependence of
 $\overline{\mathcal{F}}$ as a function of $|\lambda - \lambda_c^\alpha|$, on a log-log scale, 
 for $N= 400$ and $\alpha = 0.2$ and $0.4$. The normalized long-time averaged value of the OTOC decreases 
 following a  power law when $\lambda$ tends to $\lambda_c^\alpha$,  
 $\overline{\mathcal{F}}\propto|\lambda-\lambda_c^\alpha|^{\gamma_\lambda}$, 
 regardless of the value of $\alpha$ or $N$. By employing a least-squares fit, 
 we find that for a different control parameter and system size values, $\gamma_\lambda\approx0.36(1)$,  
 which differs from the exponent obtained for the ground-state QPT order parameter \cite{Botet1983}.
 The aforementioned results confirm that the presence of an ESQPT in the LMG system is clearly revealed by 
 the long-time averaged value of the OTOC and indicate that  $\overline{F}$ 
 may play the role of an order parameter for the ESQPT.

 So far, we have discussed the possibility of using an OTOC to probe the ESQPT in the LMG
 model parameter space, verifying that the long-time averaged value of
 the OTOC behaves as an order parameter for the ESQPT.
 However, as already mentioned, it is well known that ESQPTs are characterized by a
 singularity in the density of states at the critical energy in the mean-field limit.  In order to
 further confirm our claim that the OTOC may be an order parameter for the ESQPT in the LMG model, 
 we proceed further by studying the OTOC time dependence for different energy levels.

 To this end, we investigate the properties of the microcanonical OTOC \cite{Hashimoto2017,CCarlos2019}, defined as
 \be
   F_n(t)=\la n|{W}^\dag(t){V}^\dag(0){W}(t){V}(0)|n\ra~,
 \ee
 where, as previously, ${V}={W}=S_x/\mathcal{S}$ and
 $|n\ra$ is the $n$th eigenstate of the LMG Hamiltonian (\ref{LMG}), whose energy is $E_n$.
 We fix the parameter $\alpha$, compute the Heisenberg representation of  ${W}$,
 ${W}(t)=e^{iHt}{W}e^{-iHt}$, and calculate the time dependence of $F_n(t)$
 for different energy levels.

 
 We plot in Fig.~\ref{Etmc} the time dependence of $F_n(t)$ for three
 representative energies of the system: one that lies below the critical
 energy [Fig.~\ref{Etmc}(a)], the closest level to the critical energy level [Fig.~\ref{Etmc}(b)], 
 and a third one, well above  the critical energy [Fig.~\ref{Etmc}(c)]. 
 From this figure one can see that the imaginary part of $F_n(t)$ is always zero, or close to zero, in the three
 cases considered. For the ground state $n=0$ [cf.~Fig.~\ref{Etmc}(a)], the real part of $F_n(t)$ 
 oscillates around a finite value with a small amplitude [see the inset of Fig.~\ref{Etmc}(a)].
 Increasing the energy level $n$ gradually makes the real part of $F_n(t)$ oscillate around 
 a smaller value, simultaneously increasing the oscillation amplitude. In the case of the energy level closest to  
 the critical energy, both the imaginary and real parts of $F_n(t)$ oscillate around zero 
 with irregular patterns [cf.~Fig.~\ref{Etmc}(b)]. Once $E>E_c$, the real part of $F_n(t)$, $F_{n,R}(t)$,
 exhibits small oscillations around zero, of an amplitude that decreases with time. 
 This indicates that the steady-state value of $F_n(t)$ is zero.    
 Therefore, the long-time averaged value of $F_n(t)$ is close to zero in the neighborhood
 of the critical energy.

 The $F_n(t)$ features shown in the  three panels of Fig.~\ref{Etmc} imply a drastic change of 
 the dynamical properties of the microcanonical OTOC at $E = E_c = 0$, something that is consistent with
 the role played by the ESQPT critical energy in the LMG model.
 Hence, the existence of an ESQPT at the critical energy $E_c=0$ in the LMG model can
 be unambiguously detected from the microcanonical OTOC dynamics.
 Moreover, one can also expect that states belonging to different phases of the ESQPT
 may be characterized, in energy space, by different long-time averaged values. 
 Therefore, in a similar way as in Eq.~(\ref{AvgF}), we compute a long-time averaged value of $F_{n,R}(t)$.
 To check this we calculate the steady-state value of $F_n(t)$,  denoted by $\overline{F}_n$,
 over a time interval $T=10^4$. The obtained results are plotted in Fig.~\ref{otceng}, 
 where we plot $\overline{\mathcal{F}}_n$, defined  normalizing $\overline{F}_n$ by $\overline{F}_0$, 
 $\overline{\mathcal{F}}_n=\overline{F}_n/\overline{F}_0$, as a function of
 the rescaled system energy levels $\tilde{\varepsilon}_n=2(E_n-E_0)/(E_{max}-E_0)$.
 Here $E_0$ is the  ground-state energy and $E_{max}$ denotes the maximum energy value.
 
From Fig.~\ref{otceng}(a) one can be aware that $\overline{\mathcal{F}}_n$ is nonzero for 
$\tilde{\varepsilon}_n<\tilde{\varepsilon}^\alpha_c$,
 vanishing gradually as the energy of the system approaches the
 ESQPT critical energy. This is akin to the behavior exhibited by a ground-state QPT 
 standard order parameter \cite{Botet1983}.  
 The transition point, at which the value of
 $\overline{\mathcal{F}}_n$ becomes zero, is in the vicinity of the rescaled critical
 energy $\tilde{\varepsilon}^\alpha_c$ as $N\to\infty$. 
 It is worth emphasizing that  $\overline{\mathcal{F}}_n$ undergoes an abrupt change around the critical energy. 
 This can be explained as an effect of the system's finite size. 
 In fact, finite-size systems have a set of eigenstates with energies close to the critical one. 
 Hence, the value of $\overline{\mathcal{F}}_n$ for these eigenstates will be very small. 
 Outside this region, the eigenstates with $\tilde{\varepsilon}_n<\tilde{\varepsilon}^\alpha_c$ 
 will lead to finite $\overline{\mathcal{F}}_n$ values and at the boundary of this region 
 $\overline{\mathcal{F}}_n$ exhibits an abrupt change. 
The width of this region decreases as $N$ increases, and one can expect that the sudden change amplitude of
 $\overline{\mathcal{F}}_n$ in the neighborhood of the critical point will eventually disappear in the thermodynamic limit. 
 The inset of Fig.~\ref{otceng}(a) shows how the sudden change in the vicinity of the critical energy, which we denoted by 
 $\mathcal{D}_n$, changes with the system size $N$. 
 Here $\mathcal{D}_n$ is defined as the difference between the maximum and minimum values of 
 $\overline{\mathcal{F}}_n$ in the region $n\in[n_c-15, n_c+5]$, 
 where $n_c$ is the number of energy levels with $E_{n_c}\approx E_c=0$. 
 We have carefully checked that the abrupt change in $\overline{\mathcal{F}}_n$ is indeed located in the above region.
 Clearly, for different values of $\alpha$, $\mathcal{D}_n$ 
 quickly decreases and approaches zero for increasing $N$.
 The above-mentioned behavior of $\overline{\mathcal{F}}_n$ 
 can be clearly observed for other $\alpha$ values different from $\alpha = 0.4$, 
 for a given system size [cf.~Fig.~\ref{otceng}(b)].
 Furthermore, the inset in Fig.~\ref{otceng}(b) shows that 
 the decay of $\overline{\mathcal{F}}_n$  close to the critical energy also follows a power law, i.e., 
 $\overline{\mathcal{F}}_n\propto|\tilde{\varepsilon}_n-\tilde{\varepsilon}_c^\alpha|^{\gamma_\varepsilon}$,
 with $\gamma_\varepsilon\approx0.69(5)$, irrespective of the $\alpha$ value.

The results shown in the above figures support our claim that the ESQPT in the LMG model 
has a strong impact on the OTOC dynamics.  
The ESQPT existence can be clearly detected
 from the OTOC time dependence and its dynamical behavior. In
 particular, the OTOC long-time averaged value can be used as an order
 parameter for the ESQPT.

\section {Conclusion}\label{CSF}

 To summarize, we have investigated in detail how
 an OTOC in a two-level quantum many-body
 system (the LMG model) is affected by the existence of an ESQPT, 
 characterized by the divergence in the local density of states.  
 We have shown that the ESQPT has significant effects on the OTOC
 dynamics.  As a consequence, the calculation of the OTOC dynamics for the system eigenstates allowed us to detect 
 the presence of an ESQPT in the spectrum, estimate the critical energy value for this ESQPT, and distinguish between the 
 different phases.
 
 The definition of an order parameter for ESQPTs like the one presented in this work 
 is still an open problem \cite{Caprio2008} which is the subject of current studies \cite{PPerez2017}.  
 The connection between the OTOC
 and the ESQPT that we have presented  provides a possible venue to define an
 order parameter for ESQPTs, offering a deeper understanding on these quantum phase transitions.  
 We should point out that, even though the ESQPT studied in this work is a particular one, 
 characterized by a divergence of the density of states at the critical energy in the thermodynamic limit, 
 we expect that our results are still valid
 for other ESQPT types, e.g., those ESQPTs identified by a divergence in the first-order
 derivative of the density of states with respect to the energy.  
 Another relevant issue is the identification of the requirements that operators $W$ and $V$ should fulfill in order that the OTOC can be used as a detector between the different quantum phases.
Very recent semianalytical results \cite{Dag2019} imply that
to distinguish phases in many-body systems through OTOCs, $W$ and $V$ should be chosen as the order parameter of the equilibrium phase transition. However, in order to get a deeper perspective and to reach more general conclusions on the relation 
between OTOCs and quantum phase transitions, further work is still needed.
We hope that our present results encourage others to
explore the dynamic signatures of ESQPTs via OTOCs in the future.


 Finally, the successful measurement of OTOCs  in recent experiments involving different quantum many-body systems
 \cite{Li2017,Wei2018,Garttner2017} may pave the way to an experimental test of the obtained results in quantum simulators.  This provides a promising venue for both theoretical and experimental investigation of ESQPTs in quantum interacting systems.

 \acknowledgments 
 
 Q.W. is very grateful to
 the the Center for Theoretical Physics of Complex Systems,
 Institute for Basic Science, South Korea.
 This work was supported by National Natural Science Foundation of China (Grant No.~11805165). F.P.~B. thanks the
  Consejería de Conocimiento, Investigación y Universidad, Junta de Andalucía,
  and European Regional Development Fund (Grant No. SOMM17/6105/UGR), and
  the Centro de Estudios Avanzados de Física, Matemáticas y Computación
  (CEAFMC) of the Universidad de Huelva. Computer resources supporting
  this work were provided in part by the CEAFMC and Universidad de Huelva High
  Performance Computer located at the Campus Universitario
  el Carmen and funded by FEDER/MINECO Project No.~UNHU-15CE-2848.

\bibliographystyle{apsrev4-1}
\bibliography{OTOCb}

\begin{thebibliography}{78}%
\makeatletter
\providecommand \@ifxundefined [1]{%
 \@ifx{#1\undefined}
}%
\providecommand \@ifnum [1]{%
 \ifnum #1\expandafter \@firstoftwo
 \else \expandafter \@secondoftwo
 \fi
}%
\providecommand \@ifx [1]{%
 \ifx #1\expandafter \@firstoftwo
 \else \expandafter \@secondoftwo
 \fi
}%
\providecommand \natexlab [1]{#1}%
\providecommand \enquote  [1]{``#1''}%
\providecommand \bibnamefont  [1]{#1}%
\providecommand \bibfnamefont [1]{#1}%
\providecommand \citenamefont [1]{#1}%
\providecommand \href@noop [0]{\@secondoftwo}%
\providecommand \href [0]{\begingroup \@sanitize@url \@href}%
\providecommand \@href[1]{\@@startlink{#1}\@@href}%
\providecommand \@@href[1]{\endgroup#1\@@endlink}%
\providecommand \@sanitize@url [0]{\catcode `\\12\catcode `\$12\catcode
  `\&12\catcode `\#12\catcode `\^12\catcode `\_12\catcode `\%12\relax}%
\providecommand \@@startlink[1]{}%
\providecommand \@@endlink[0]{}%
\providecommand \url  [0]{\begingroup\@sanitize@url \@url }%
\providecommand \@url [1]{\endgroup\@href {#1}{\urlprefix }}%
\providecommand \urlprefix  [0]{URL }%
\providecommand \Eprint [0]{\href }%
\providecommand \doibase [0]{http://dx.doi.org/}%
\providecommand \selectlanguage [0]{\@gobble}%
\providecommand \bibinfo  [0]{\@secondoftwo}%
\providecommand \bibfield  [0]{\@secondoftwo}%
\providecommand \translation [1]{[#1]}%
\providecommand \BibitemOpen [0]{}%
\providecommand \bibitemStop [0]{}%
\providecommand \bibitemNoStop [0]{.\EOS\space}%
\providecommand \EOS [0]{\spacefactor3000\relax}%
\providecommand \BibitemShut  [1]{\csname bibitem#1\endcsname}%
\let\auto@bib@innerbib\@empty
\bibitem [{\citenamefont {Carr}(2010)}]{Carr2010}%
  \BibitemOpen
  \bibfield  {author} {\bibinfo {author} {\bibfnamefont {L.}~\bibnamefont
  {Carr}},\ }\href@noop {} {\emph {\bibinfo {title} {Understanding Quantum
  Phase Transitions}}},\ Condensed Matter Physics\ (\bibinfo  {publisher}
  {Taylor and Francis},\ \bibinfo {address} {Hoboken},\ \bibinfo {year}
  {2010})\BibitemShut {NoStop}%
\bibitem [{\citenamefont {Cejnar}\ \emph {et~al.}(2006)\citenamefont {Cejnar},
  \citenamefont {Macek}, \citenamefont {Heinze}, \citenamefont {Jolie},\ and\
  \citenamefont {Dobes}}]{Cejnar2006}%
  \BibitemOpen
  \bibfield  {author} {\bibinfo {author} {\bibfnamefont {P.}~\bibnamefont
  {Cejnar}}, \bibinfo {author} {\bibfnamefont {M.}~\bibnamefont {Macek}},
  \bibinfo {author} {\bibfnamefont {S.}~\bibnamefont {Heinze}}, \bibinfo
  {author} {\bibfnamefont {J.}~\bibnamefont {Jolie}}, \ and\ \bibinfo {author}
  {\bibfnamefont {J.}~\bibnamefont {Dobes}},\ }\href@noop {} {\bibfield
  {journal} {\bibinfo  {journal} {J. Phys. A: Math. and Gen.}\ }\textbf
  {\bibinfo {volume} {39}},\ \bibinfo {pages} {L515} (\bibinfo {year}
  {2006})}\BibitemShut {NoStop}%
\bibitem [{\citenamefont {Caprio}\ \emph {et~al.}(2008)\citenamefont {Caprio},
  \citenamefont {Cejnar},\ and\ \citenamefont {Iachello}}]{Caprio2008}%
  \BibitemOpen
  \bibfield  {author} {\bibinfo {author} {\bibfnamefont {M.}~\bibnamefont
  {Caprio}}, \bibinfo {author} {\bibfnamefont {P.}~\bibnamefont {Cejnar}}, \
  and\ \bibinfo {author} {\bibfnamefont {F.}~\bibnamefont {Iachello}},\
  }\href@noop {} {\bibfield  {journal} {\bibinfo  {journal} {Ann. Phys.}\
  }\textbf {\bibinfo {volume} {323}},\ \bibinfo {pages} {1106 } (\bibinfo
  {year} {2008})}\BibitemShut {NoStop}%
\bibitem [{\citenamefont {Brandes}(2013)}]{Brandes2013}%
  \BibitemOpen
  \bibfield  {author} {\bibinfo {author} {\bibfnamefont {T.}~\bibnamefont
  {Brandes}},\ }\href@noop {} {\bibfield  {journal} {\bibinfo  {journal} {Phys.
  Rev. E}\ }\textbf {\bibinfo {volume} {88}},\ \bibinfo {pages} {032133}
  (\bibinfo {year} {2013})}\BibitemShut {NoStop}%
\bibitem [{\citenamefont {Str\'ansk\'y}\ \emph {et~al.}(2014)\citenamefont
  {Str\'ansk\'y}, \citenamefont {Macek},\ and\ \citenamefont
  {Cejnar}}]{Stransky2014}%
  \BibitemOpen
  \bibfield  {author} {\bibinfo {author} {\bibfnamefont {P.}~\bibnamefont
  {Str\'ansk\'y}}, \bibinfo {author} {\bibfnamefont {M.}~\bibnamefont {Macek}},
  \ and\ \bibinfo {author} {\bibfnamefont {P.}~\bibnamefont {Cejnar}},\
  }\href@noop {} {\bibfield  {journal} {\bibinfo  {journal} {Ann. Phys.}\
  }\textbf {\bibinfo {volume} {345}},\ \bibinfo {pages} {73 } (\bibinfo {year}
  {2014})}\BibitemShut {NoStop}%
\bibitem [{\citenamefont {Santos}\ \emph {et~al.}(2016)\citenamefont {Santos},
  \citenamefont {T\'avora},\ and\ \citenamefont {P\'erez-Bernal}}]{Santos2016}%
  \BibitemOpen
  \bibfield  {author} {\bibinfo {author} {\bibfnamefont {L.~F.}\ \bibnamefont
  {Santos}}, \bibinfo {author} {\bibfnamefont {M.}~\bibnamefont {T\'avora}}, \
  and\ \bibinfo {author} {\bibfnamefont {F.}~\bibnamefont {P\'erez-Bernal}},\
  }\href@noop {} {\bibfield  {journal} {\bibinfo  {journal} {Phys. Rev. A}\
  }\textbf {\bibinfo {volume} {94}},\ \bibinfo {pages} {012113} (\bibinfo
  {year} {2016})}\BibitemShut {NoStop}%
\bibitem [{\citenamefont {Santos}\ and\ \citenamefont
  {P\'erez-Bernal}(2015)}]{Santos2015}%
  \BibitemOpen
  \bibfield  {author} {\bibinfo {author} {\bibfnamefont {L.~F.}\ \bibnamefont
  {Santos}}\ and\ \bibinfo {author} {\bibfnamefont {F.}~\bibnamefont
  {P\'erez-Bernal}},\ }\href@noop {} {\bibfield  {journal} {\bibinfo  {journal}
  {Phys. Rev. A}\ }\textbf {\bibinfo {volume} {92}},\ \bibinfo {pages} {050101}
  (\bibinfo {year} {2015})}\BibitemShut {NoStop}%
\bibitem [{\citenamefont {P\'erez-Bernal}\ and\ \citenamefont
  {Santos}(2017)}]{PBernal2017}%
  \BibitemOpen
  \bibfield  {author} {\bibinfo {author} {\bibfnamefont {F.}~\bibnamefont
  {P\'erez-Bernal}}\ and\ \bibinfo {author} {\bibfnamefont {L.~F.}\
  \bibnamefont {Santos}},\ }\href {\doibase 10.1002/prop.201600035} {\bibfield
  {journal} {\bibinfo  {journal} {Progr. Phys. Fortschr. Phys.}\ }\textbf
  {\bibinfo {volume} {65}},\ \bibinfo {pages} {1600035} (\bibinfo {year}
  {2017})}\BibitemShut {NoStop}%
\bibitem [{\citenamefont {Larese}\ and\ \citenamefont
  {Iachello}(2011)}]{Larese2011}%
  \BibitemOpen
  \bibfield  {author} {\bibinfo {author} {\bibfnamefont {D.}~\bibnamefont
  {Larese}}\ and\ \bibinfo {author} {\bibfnamefont {F.}~\bibnamefont
  {Iachello}},\ }\href@noop {} {\bibfield  {journal} {\bibinfo  {journal} {J.
  Molec. Struct.}\ }\textbf {\bibinfo {volume} {1006}},\ \bibinfo {pages} {611
  } (\bibinfo {year} {2011})}\BibitemShut {NoStop}%
\bibitem [{\citenamefont {Larese}\ \emph {et~al.}(2013)\citenamefont {Larese},
  \citenamefont {P\'erez-Bernal},\ and\ \citenamefont {Iachello}}]{Larese2013}%
  \BibitemOpen
  \bibfield  {author} {\bibinfo {author} {\bibfnamefont {D.}~\bibnamefont
  {Larese}}, \bibinfo {author} {\bibfnamefont {F.}~\bibnamefont
  {P\'erez-Bernal}}, \ and\ \bibinfo {author} {\bibfnamefont {F.}~\bibnamefont
  {Iachello}},\ }\href@noop {} {\bibfield  {journal} {\bibinfo  {journal} {J.
  Molec. Struct.}\ }\textbf {\bibinfo {volume} {1051}},\ \bibinfo {pages} {310
  } (\bibinfo {year} {2013})}\BibitemShut {NoStop}%
\bibitem [{\citenamefont {Dietz}\ \emph {et~al.}(2013)\citenamefont {Dietz},
  \citenamefont {Iachello}, \citenamefont {Miski-Oglu}, \citenamefont
  {Pietralla}, \citenamefont {Richter}, \citenamefont {von Smekal},\ and\
  \citenamefont {Wambach}}]{Dietz2013}%
  \BibitemOpen
  \bibfield  {author} {\bibinfo {author} {\bibfnamefont {B.}~\bibnamefont
  {Dietz}}, \bibinfo {author} {\bibfnamefont {F.}~\bibnamefont {Iachello}},
  \bibinfo {author} {\bibfnamefont {M.}~\bibnamefont {Miski-Oglu}}, \bibinfo
  {author} {\bibfnamefont {N.}~\bibnamefont {Pietralla}}, \bibinfo {author}
  {\bibfnamefont {A.}~\bibnamefont {Richter}}, \bibinfo {author} {\bibfnamefont
  {L.}~\bibnamefont {von Smekal}}, \ and\ \bibinfo {author} {\bibfnamefont
  {J.}~\bibnamefont {Wambach}},\ }\href@noop {} {\bibfield  {journal} {\bibinfo
   {journal} {Phys. Rev. B}\ }\textbf {\bibinfo {volume} {88}},\ \bibinfo
  {pages} {104101} (\bibinfo {year} {2013})}\BibitemShut {NoStop}%
\bibitem [{\citenamefont {Zhao}\ \emph {et~al.}(2014)\citenamefont {Zhao},
  \citenamefont {Jiang}, \citenamefont {Tang}, \citenamefont {Webb},\ and\
  \citenamefont {Liu}}]{Zhao2014}%
  \BibitemOpen
  \bibfield  {author} {\bibinfo {author} {\bibfnamefont {L.}~\bibnamefont
  {Zhao}}, \bibinfo {author} {\bibfnamefont {J.}~\bibnamefont {Jiang}},
  \bibinfo {author} {\bibfnamefont {T.}~\bibnamefont {Tang}}, \bibinfo {author}
  {\bibfnamefont {M.}~\bibnamefont {Webb}}, \ and\ \bibinfo {author}
  {\bibfnamefont {Y.}~\bibnamefont {Liu}},\ }\href@noop {} {\bibfield
  {journal} {\bibinfo  {journal} {Phys. Rev. A}\ }\textbf {\bibinfo {volume}
  {89}},\ \bibinfo {pages} {023608} (\bibinfo {year} {2014})}\BibitemShut
  {NoStop}%
\bibitem [{\citenamefont {Khalouf-Rivera}\ \emph {et~al.}(2019)\citenamefont
  {Khalouf-Rivera}, \citenamefont {Carvajal}, \citenamefont {Santos},\ and\
  \citenamefont {Pérez-Bernal}}]{KRivera2019}%
  \BibitemOpen
  \bibfield  {author} {\bibinfo {author} {\bibfnamefont {J.}~\bibnamefont
  {Khalouf-Rivera}}, \bibinfo {author} {\bibfnamefont {M.}~\bibnamefont
  {Carvajal}}, \bibinfo {author} {\bibfnamefont {L.~F.}\ \bibnamefont
  {Santos}}, \ and\ \bibinfo {author} {\bibfnamefont {F.}~\bibnamefont
  {Pérez-Bernal}},\ }\href {\doibase 10.1021/acs.jpca.9b07338} {\bibfield
  {journal} {\bibinfo  {journal} {The Journal of Physical Chemistry A}\
  }\textbf {\bibinfo {volume} {123}},\ \bibinfo {pages} {9544} (\bibinfo {year}
  {2019})},\ \bibinfo {note} {pMID: 31596086},\ \Eprint
  {http://arxiv.org/abs/https://doi.org/10.1021/acs.jpca.9b07338}
  {https://doi.org/10.1021/acs.jpca.9b07338} \BibitemShut {NoStop}%
\bibitem [{\citenamefont {Puebla}\ \emph {et~al.}(2013)\citenamefont {Puebla},
  \citenamefont {Rela\~no},\ and\ \citenamefont {Retamosa}}]{Puebla2013}%
  \BibitemOpen
  \bibfield  {author} {\bibinfo {author} {\bibfnamefont {R.}~\bibnamefont
  {Puebla}}, \bibinfo {author} {\bibfnamefont {A.}~\bibnamefont {Rela\~no}}, \
  and\ \bibinfo {author} {\bibfnamefont {J.}~\bibnamefont {Retamosa}},\
  }\href@noop {} {\bibfield  {journal} {\bibinfo  {journal} {Phys. Rev. A}\
  }\textbf {\bibinfo {volume} {87}},\ \bibinfo {pages} {023819} (\bibinfo
  {year} {2013})}\BibitemShut {NoStop}%
\bibitem [{\citenamefont {Puebla}\ and\ \citenamefont
  {Rela\~no}(2015)}]{Puebla2015}%
  \BibitemOpen
  \bibfield  {author} {\bibinfo {author} {\bibfnamefont {R.}~\bibnamefont
  {Puebla}}\ and\ \bibinfo {author} {\bibfnamefont {A.}~\bibnamefont
  {Rela\~no}},\ }\href@noop {} {\bibfield  {journal} {\bibinfo  {journal}
  {Phys. Rev. E}\ }\textbf {\bibinfo {volume} {92}},\ \bibinfo {pages} {012101}
  (\bibinfo {year} {2015})}\BibitemShut {NoStop}%
\bibitem [{\citenamefont {Engelhardt}\ \emph {et~al.}(2015)\citenamefont
  {Engelhardt}, \citenamefont {Bastidas}, \citenamefont {Kopylov},\ and\
  \citenamefont {Brandes}}]{Engelhardt2015}%
  \BibitemOpen
  \bibfield  {author} {\bibinfo {author} {\bibfnamefont {G.}~\bibnamefont
  {Engelhardt}}, \bibinfo {author} {\bibfnamefont {V.~M.}\ \bibnamefont
  {Bastidas}}, \bibinfo {author} {\bibfnamefont {W.}~\bibnamefont {Kopylov}}, \
  and\ \bibinfo {author} {\bibfnamefont {T.}~\bibnamefont {Brandes}},\
  }\href@noop {} {\bibfield  {journal} {\bibinfo  {journal} {Phys. Rev. A}\
  }\textbf {\bibinfo {volume} {91}},\ \bibinfo {pages} {013631} (\bibinfo
  {year} {2015})}\BibitemShut {NoStop}%
\bibitem [{\citenamefont {Kopylov}\ and\ \citenamefont
  {Brandes}(2015)}]{Kopylov2015}%
  \BibitemOpen
  \bibfield  {author} {\bibinfo {author} {\bibfnamefont {W.}~\bibnamefont
  {Kopylov}}\ and\ \bibinfo {author} {\bibfnamefont {T.}~\bibnamefont
  {Brandes}},\ }\href@noop {} {\bibfield  {journal} {\bibinfo  {journal} {New
  J. Phys.}\ }\textbf {\bibinfo {volume} {17}},\ \bibinfo {pages} {103031}
  (\bibinfo {year} {2015})}\BibitemShut {NoStop}%
\bibitem [{\citenamefont {Kloc}\ \emph {et~al.}(2018)\citenamefont {Kloc},
  \citenamefont {Str\'ansk\'y},\ and\ \citenamefont {Cejnar}}]{Kloc2018}%
  \BibitemOpen
  \bibfield  {author} {\bibinfo {author} {\bibfnamefont {M.}~\bibnamefont
  {Kloc}}, \bibinfo {author} {\bibfnamefont {P.}~\bibnamefont {Str\'ansk\'y}},
  \ and\ \bibinfo {author} {\bibfnamefont {P.}~\bibnamefont {Cejnar}},\
  }\href@noop {} {\bibfield  {journal} {\bibinfo  {journal} {Phys. Rev. A}\
  }\textbf {\bibinfo {volume} {98}},\ \bibinfo {pages} {013836} (\bibinfo
  {year} {2018})}\BibitemShut {NoStop}%
\bibitem [{\citenamefont {Wang}\ and\ \citenamefont
  {P\'erez-Bernal}(2019)}]{Wang2019}%
  \BibitemOpen
  \bibfield  {author} {\bibinfo {author} {\bibfnamefont {Q.}~\bibnamefont
  {Wang}}\ and\ \bibinfo {author} {\bibfnamefont {F.}~\bibnamefont
  {P\'erez-Bernal}},\ }\href {\doibase 10.1103/PhysRevA.100.022118} {\bibfield
  {journal} {\bibinfo  {journal} {Phys. Rev. A}\ }\textbf {\bibinfo {volume}
  {100}},\ \bibinfo {pages} {022118} (\bibinfo {year} {2019})}\BibitemShut
  {NoStop}%
\bibitem [{\citenamefont {Rela\~no}\ \emph {et~al.}(2008)\citenamefont
  {Rela\~no}, \citenamefont {Arias}, \citenamefont {Dukelsky}, \citenamefont
  {Garc\'{\i}a-Ramos},\ and\ \citenamefont {P\'erez-Fern\'andez}}]{Relano2008}%
  \BibitemOpen
  \bibfield  {author} {\bibinfo {author} {\bibfnamefont {A.}~\bibnamefont
  {Rela\~no}}, \bibinfo {author} {\bibfnamefont {J.~M.}\ \bibnamefont {Arias}},
  \bibinfo {author} {\bibfnamefont {J.}~\bibnamefont {Dukelsky}}, \bibinfo
  {author} {\bibfnamefont {J.~E.}\ \bibnamefont {Garc\'{\i}a-Ramos}}, \ and\
  \bibinfo {author} {\bibfnamefont {P.}~\bibnamefont {P\'erez-Fern\'andez}},\
  }\href@noop {} {\bibfield  {journal} {\bibinfo  {journal} {Phys. Rev. A}\
  }\textbf {\bibinfo {volume} {78}},\ \bibinfo {pages} {060102} (\bibinfo
  {year} {2008})}\BibitemShut {NoStop}%
\bibitem [{\citenamefont {P\'erez-Fern\'andez}\ \emph
  {et~al.}(2009)\citenamefont {P\'erez-Fern\'andez}, \citenamefont {Rela\~no},
  \citenamefont {Arias}, \citenamefont {Dukelsky},\ and\ \citenamefont
  {Garc\'{\i}a-Ramos}}]{PFernandez2009}%
  \BibitemOpen
  \bibfield  {author} {\bibinfo {author} {\bibfnamefont {P.}~\bibnamefont
  {P\'erez-Fern\'andez}}, \bibinfo {author} {\bibfnamefont {A.}~\bibnamefont
  {Rela\~no}}, \bibinfo {author} {\bibfnamefont {J.~M.}\ \bibnamefont {Arias}},
  \bibinfo {author} {\bibfnamefont {J.}~\bibnamefont {Dukelsky}}, \ and\
  \bibinfo {author} {\bibfnamefont {J.~E.}\ \bibnamefont {Garc\'{\i}a-Ramos}},\
  }\href@noop {} {\bibfield  {journal} {\bibinfo  {journal} {Phys. Rev. A}\
  }\textbf {\bibinfo {volume} {80}},\ \bibinfo {pages} {032111} (\bibinfo
  {year} {2009})}\BibitemShut {NoStop}%
\bibitem [{\citenamefont {Wang}\ and\ \citenamefont {Quan}(2017)}]{Wang2017}%
  \BibitemOpen
  \bibfield  {author} {\bibinfo {author} {\bibfnamefont {Q.}~\bibnamefont
  {Wang}}\ and\ \bibinfo {author} {\bibfnamefont {H.~T.}\ \bibnamefont
  {Quan}},\ }\href@noop {} {\bibfield  {journal} {\bibinfo  {journal} {Phys.
  Rev. E}\ }\textbf {\bibinfo {volume} {96}},\ \bibinfo {pages} {032142}
  (\bibinfo {year} {2017})}\BibitemShut {NoStop}%
\bibitem [{\citenamefont {L\'obez}\ and\ \citenamefont
  {Rela\~no}(2016)}]{Lobez2016}%
  \BibitemOpen
  \bibfield  {author} {\bibinfo {author} {\bibfnamefont {C.~M.}\ \bibnamefont
  {L\'obez}}\ and\ \bibinfo {author} {\bibfnamefont {A.}~\bibnamefont
  {Rela\~no}},\ }\href@noop {} {\bibfield  {journal} {\bibinfo  {journal}
  {Phys. Rev. E}\ }\textbf {\bibinfo {volume} {94}},\ \bibinfo {pages} {012140}
  (\bibinfo {year} {2016})}\BibitemShut {NoStop}%
\bibitem [{\citenamefont {Larkin}\ and\ \citenamefont
  {Ovchinnikov}(1969)}]{Larkin1969}%
  \BibitemOpen
  \bibfield  {author} {\bibinfo {author} {\bibfnamefont {A.~I.}\ \bibnamefont
  {Larkin}}\ and\ \bibinfo {author} {\bibfnamefont {Y.~N.}\ \bibnamefont
  {Ovchinnikov}},\ }\href@noop {} {\bibfield  {journal} {\bibinfo  {journal}
  {Sov. Phys. JETP}\ }\textbf {\bibinfo {volume} {28}},\ \bibinfo {pages}
  {1200} (\bibinfo {year} {1969})}\BibitemShut {NoStop}%
\bibitem [{\citenamefont {Kitaev}(2015)}]{kitaev}%
  \BibitemOpen
  \bibfield  {author} {\bibinfo {author} {\bibfnamefont {A.}~\bibnamefont
  {Kitaev}},\ }\href@noop {} {\enquote {\bibinfo {title} {A simple model of
  quantum holography},}\ }\bibinfo {howpublished}
  {\url{http://online.kitp.ucsb.edu/online/entangled15/kitaev/}} (\bibinfo
  {year} {2015}),\ \bibinfo {note} {{KITP} Program: Entanglement in Strongly
  Correlated Quantum Matter}\BibitemShut {NoStop}%
\bibitem [{\citenamefont {Swingle}(2018)}]{Swingle2018}%
  \BibitemOpen
  \bibfield  {author} {\bibinfo {author} {\bibfnamefont {B.}~\bibnamefont
  {Swingle}},\ }\href {\doibase 10.1038/s41567-018-0295-5} {\bibfield
  {journal} {\bibinfo  {journal} {Nat. Phys.}\ }\textbf {\bibinfo {volume}
  {14}},\ \bibinfo {pages} {988} (\bibinfo {year} {2018})}\BibitemShut
  {NoStop}%
\bibitem [{\citenamefont {Roberts}\ and\ \citenamefont
  {Swingle}(2016)}]{Roberts2016}%
  \BibitemOpen
  \bibfield  {author} {\bibinfo {author} {\bibfnamefont {D.~A.}\ \bibnamefont
  {Roberts}}\ and\ \bibinfo {author} {\bibfnamefont {B.}~\bibnamefont
  {Swingle}},\ }\href {\doibase 10.1103/PhysRevLett.117.091602} {\bibfield
  {journal} {\bibinfo  {journal} {Phys. Rev. Lett.}\ }\textbf {\bibinfo
  {volume} {117}},\ \bibinfo {pages} {091602} (\bibinfo {year}
  {2016})}\BibitemShut {NoStop}%
\bibitem [{\citenamefont {Luitz}\ and\ \citenamefont
  {Bar~Lev}(2017)}]{Luitz2017}%
  \BibitemOpen
  \bibfield  {author} {\bibinfo {author} {\bibfnamefont {D.~J.}\ \bibnamefont
  {Luitz}}\ and\ \bibinfo {author} {\bibfnamefont {Y.}~\bibnamefont
  {Bar~Lev}},\ }\href {\doibase 10.1103/PhysRevB.96.020406} {\bibfield
  {journal} {\bibinfo  {journal} {Phys. Rev. B}\ }\textbf {\bibinfo {volume}
  {96}},\ \bibinfo {pages} {020406} (\bibinfo {year} {2017})}\BibitemShut
  {NoStop}%
\bibitem [{\citenamefont {Iyoda}\ and\ \citenamefont
  {Sagawa}(2018)}]{Iyoda2018}%
  \BibitemOpen
  \bibfield  {author} {\bibinfo {author} {\bibfnamefont {E.}~\bibnamefont
  {Iyoda}}\ and\ \bibinfo {author} {\bibfnamefont {T.}~\bibnamefont {Sagawa}},\
  }\href {\doibase 10.1103/PhysRevA.97.042330} {\bibfield  {journal} {\bibinfo
  {journal} {Phys. Rev. A}\ }\textbf {\bibinfo {volume} {97}},\ \bibinfo
  {pages} {042330} (\bibinfo {year} {2018})}\BibitemShut {NoStop}%
\bibitem [{\citenamefont {von Keyserlingk}\ \emph {et~al.}(2018)\citenamefont
  {von Keyserlingk}, \citenamefont {Rakovszky}, \citenamefont {Pollmann},\ and\
  \citenamefont {Sondhi}}]{Keyserlingk2018}%
  \BibitemOpen
  \bibfield  {author} {\bibinfo {author} {\bibfnamefont {C.~W.}\ \bibnamefont
  {von Keyserlingk}}, \bibinfo {author} {\bibfnamefont {T.}~\bibnamefont
  {Rakovszky}}, \bibinfo {author} {\bibfnamefont {F.}~\bibnamefont {Pollmann}},
  \ and\ \bibinfo {author} {\bibfnamefont {S.~L.}\ \bibnamefont {Sondhi}},\
  }\href {\doibase 10.1103/PhysRevX.8.021013} {\bibfield  {journal} {\bibinfo
  {journal} {Phys. Rev. X}\ }\textbf {\bibinfo {volume} {8}},\ \bibinfo {pages}
  {021013} (\bibinfo {year} {2018})}\BibitemShut {NoStop}%
\bibitem [{\citenamefont {{Niknam}}\ \emph {et~al.}(2018)\citenamefont
  {{Niknam}}, \citenamefont {{Santos}},\ and\ \citenamefont
  {{Cory}}}]{Niknam2018}%
  \BibitemOpen
  \bibfield  {author} {\bibinfo {author} {\bibfnamefont {M.}~\bibnamefont
  {{Niknam}}}, \bibinfo {author} {\bibfnamefont {L.~F.}\ \bibnamefont
  {{Santos}}}, \ and\ \bibinfo {author} {\bibfnamefont {D.~G.}\ \bibnamefont
  {{Cory}}},\ }\href@noop {} {\bibfield  {journal} {\bibinfo  {journal} {arXiv
  e-prints}\ } (\bibinfo {year} {2018})},\ \Eprint
  {http://arxiv.org/abs/1808.04375} {arXiv:1808.04375 [quant-ph]} \BibitemShut
  {NoStop}%
\bibitem [{\citenamefont {Lewis-Swan}\ \emph {et~al.}(2019)\citenamefont
  {Lewis-Swan}, \citenamefont {Safavi-Naini}, \citenamefont {Bollinger},\ and\
  \citenamefont {Rey}}]{LSwan2019}%
  \BibitemOpen
  \bibfield  {author} {\bibinfo {author} {\bibfnamefont {R.~J.}\ \bibnamefont
  {Lewis-Swan}}, \bibinfo {author} {\bibfnamefont {A.}~\bibnamefont
  {Safavi-Naini}}, \bibinfo {author} {\bibfnamefont {J.~J.}\ \bibnamefont
  {Bollinger}}, \ and\ \bibinfo {author} {\bibfnamefont {A.~M.}\ \bibnamefont
  {Rey}},\ }\href {\doibase 10.1038/s41467-019-09436-y} {\bibfield  {journal}
  {\bibinfo  {journal} {Nat. Commun.}\ }\textbf {\bibinfo {volume} {10}},\
  \bibinfo {pages} {1581} (\bibinfo {year} {2019})}\BibitemShut {NoStop}%
\bibitem [{\citenamefont {Alavirad}\ and\ \citenamefont
  {Lavasani}(2019)}]{Yahya2019}%
  \BibitemOpen
  \bibfield  {author} {\bibinfo {author} {\bibfnamefont {Y.}~\bibnamefont
  {Alavirad}}\ and\ \bibinfo {author} {\bibfnamefont {A.}~\bibnamefont
  {Lavasani}},\ }\href {\doibase 10.1103/PhysRevA.99.043602} {\bibfield
  {journal} {\bibinfo  {journal} {Phys. Rev. A}\ }\textbf {\bibinfo {volume}
  {99}},\ \bibinfo {pages} {043602} (\bibinfo {year} {2019})}\BibitemShut
  {NoStop}%
\bibitem [{\citenamefont {Maldacena}\ \emph {et~al.}(2016)\citenamefont
  {Maldacena}, \citenamefont {Shenker},\ and\ \citenamefont
  {Stanford}}]{Maldacena2016}%
  \BibitemOpen
  \bibfield  {author} {\bibinfo {author} {\bibfnamefont {J.}~\bibnamefont
  {Maldacena}}, \bibinfo {author} {\bibfnamefont {S.~H.}\ \bibnamefont
  {Shenker}}, \ and\ \bibinfo {author} {\bibfnamefont {D.}~\bibnamefont
  {Stanford}},\ }\href {\doibase 10.1007/JHEP08(2016)106} {\bibfield  {journal}
  {\bibinfo  {journal} {J. High Energy Phys.}\ }\textbf {\bibinfo {volume}
  {2016}},\ \bibinfo {pages} {106} (\bibinfo {year} {2016})}\BibitemShut
  {NoStop}%
\bibitem [{\citenamefont {Hosur}\ \emph {et~al.}(2016)\citenamefont {Hosur},
  \citenamefont {Qi}, \citenamefont {Roberts},\ and\ \citenamefont
  {Yoshida}}]{Hosur2016}%
  \BibitemOpen
  \bibfield  {author} {\bibinfo {author} {\bibfnamefont {P.}~\bibnamefont
  {Hosur}}, \bibinfo {author} {\bibfnamefont {X.-L.}\ \bibnamefont {Qi}},
  \bibinfo {author} {\bibfnamefont {D.~A.}\ \bibnamefont {Roberts}}, \ and\
  \bibinfo {author} {\bibfnamefont {B.}~\bibnamefont {Yoshida}},\ }\href
  {\doibase 10.1007/JHEP02(2016)004} {\bibfield  {journal} {\bibinfo  {journal}
  {J. High Energy Phys.}\ }\textbf {\bibinfo {volume} {2016}},\ \bibinfo
  {pages} {4} (\bibinfo {year} {2016})}\BibitemShut {NoStop}%
\bibitem [{\citenamefont {Kukuljan}\ \emph {et~al.}(2017)\citenamefont
  {Kukuljan}, \citenamefont {Grozdanov},\ and\ \citenamefont
  {Prosen}}]{Kukuljan2017}%
  \BibitemOpen
  \bibfield  {author} {\bibinfo {author} {\bibfnamefont {I.}~\bibnamefont
  {Kukuljan}}, \bibinfo {author} {\bibfnamefont {S.~c.~v.}\ \bibnamefont
  {Grozdanov}}, \ and\ \bibinfo {author} {\bibfnamefont {T.~c.~v.}\
  \bibnamefont {Prosen}},\ }\href {\doibase 10.1103/PhysRevB.96.060301}
  {\bibfield  {journal} {\bibinfo  {journal} {Phys. Rev. B}\ }\textbf {\bibinfo
  {volume} {96}},\ \bibinfo {pages} {060301} (\bibinfo {year}
  {2017})}\BibitemShut {NoStop}%
\bibitem [{\citenamefont {Hashimoto}\ \emph {et~al.}(2017)\citenamefont
  {Hashimoto}, \citenamefont {Murata},\ and\ \citenamefont
  {Yoshii}}]{Hashimoto2017}%
  \BibitemOpen
  \bibfield  {author} {\bibinfo {author} {\bibfnamefont {K.}~\bibnamefont
  {Hashimoto}}, \bibinfo {author} {\bibfnamefont {K.}~\bibnamefont {Murata}}, \
  and\ \bibinfo {author} {\bibfnamefont {R.}~\bibnamefont {Yoshii}},\ }\href
  {\doibase 10.1007/JHEP10(2017)138} {\bibfield  {journal} {\bibinfo  {journal}
  {J. High Energy Phys.}\ }\textbf {\bibinfo {volume} {2017}},\ \bibinfo
  {pages} {138} (\bibinfo {year} {2017})}\BibitemShut {NoStop}%
\bibitem [{\citenamefont {Rozenbaum}\ \emph {et~al.}(2017)\citenamefont
  {Rozenbaum}, \citenamefont {Ganeshan},\ and\ \citenamefont
  {Galitski}}]{Rozenbaum2017}%
  \BibitemOpen
  \bibfield  {author} {\bibinfo {author} {\bibfnamefont {E.~B.}\ \bibnamefont
  {Rozenbaum}}, \bibinfo {author} {\bibfnamefont {S.}~\bibnamefont {Ganeshan}},
  \ and\ \bibinfo {author} {\bibfnamefont {V.}~\bibnamefont {Galitski}},\
  }\href {\doibase 10.1103/PhysRevLett.118.086801} {\bibfield  {journal}
  {\bibinfo  {journal} {Phys. Rev. Lett.}\ }\textbf {\bibinfo {volume} {118}},\
  \bibinfo {pages} {086801} (\bibinfo {year} {2017})}\BibitemShut {NoStop}%
\bibitem [{\citenamefont {Rozenbaum}\ \emph {et~al.}(2019)\citenamefont
  {Rozenbaum}, \citenamefont {Ganeshan},\ and\ \citenamefont
  {Galitski}}]{Rozenbaum2019}%
  \BibitemOpen
  \bibfield  {author} {\bibinfo {author} {\bibfnamefont {E.~B.}\ \bibnamefont
  {Rozenbaum}}, \bibinfo {author} {\bibfnamefont {S.}~\bibnamefont {Ganeshan}},
  \ and\ \bibinfo {author} {\bibfnamefont {V.}~\bibnamefont {Galitski}},\
  }\href {\doibase 10.1103/PhysRevB.100.035112} {\bibfield  {journal} {\bibinfo
   {journal} {Phys. Rev. B}\ }\textbf {\bibinfo {volume} {100}},\ \bibinfo
  {pages} {035112} (\bibinfo {year} {2019})}\BibitemShut {NoStop}%
\bibitem [{\citenamefont {Garc\'{\i}a-Mata}\ \emph {et~al.}(2018)\citenamefont
  {Garc\'{\i}a-Mata}, \citenamefont {Saraceno}, \citenamefont {Jalabert},
  \citenamefont {Roncaglia},\ and\ \citenamefont {Wisniacki}}]{GMata2018}%
  \BibitemOpen
  \bibfield  {author} {\bibinfo {author} {\bibfnamefont {I.}~\bibnamefont
  {Garc\'{\i}a-Mata}}, \bibinfo {author} {\bibfnamefont {M.}~\bibnamefont
  {Saraceno}}, \bibinfo {author} {\bibfnamefont {R.~A.}\ \bibnamefont
  {Jalabert}}, \bibinfo {author} {\bibfnamefont {A.~J.}\ \bibnamefont
  {Roncaglia}}, \ and\ \bibinfo {author} {\bibfnamefont {D.~A.}\ \bibnamefont
  {Wisniacki}},\ }\href {\doibase 10.1103/PhysRevLett.121.210601} {\bibfield
  {journal} {\bibinfo  {journal} {Phys. Rev. Lett.}\ }\textbf {\bibinfo
  {volume} {121}},\ \bibinfo {pages} {210601} (\bibinfo {year}
  {2018})}\BibitemShut {NoStop}%
\bibitem [{\citenamefont {Jalabert}\ \emph {et~al.}(2018)\citenamefont
  {Jalabert}, \citenamefont {Garc\'{\i}a-Mata},\ and\ \citenamefont
  {Wisniacki}}]{Jalabert2018}%
  \BibitemOpen
  \bibfield  {author} {\bibinfo {author} {\bibfnamefont {R.~A.}\ \bibnamefont
  {Jalabert}}, \bibinfo {author} {\bibfnamefont {I.}~\bibnamefont
  {Garc\'{\i}a-Mata}}, \ and\ \bibinfo {author} {\bibfnamefont {D.~A.}\
  \bibnamefont {Wisniacki}},\ }\href {\doibase 10.1103/PhysRevE.98.062218}
  {\bibfield  {journal} {\bibinfo  {journal} {Phys. Rev. E}\ }\textbf {\bibinfo
  {volume} {98}},\ \bibinfo {pages} {062218} (\bibinfo {year}
  {2018})}\BibitemShut {NoStop}%
\bibitem [{\citenamefont {Fortes}\ \emph {et~al.}(2019)\citenamefont {Fortes},
  \citenamefont {Garc\'{\i}a-Mata}, \citenamefont {Jalabert},\ and\
  \citenamefont {Wisniacki}}]{Fortes2019}%
  \BibitemOpen
  \bibfield  {author} {\bibinfo {author} {\bibfnamefont {E.~M.}\ \bibnamefont
  {Fortes}}, \bibinfo {author} {\bibfnamefont {I.}~\bibnamefont
  {Garc\'{\i}a-Mata}}, \bibinfo {author} {\bibfnamefont {R.~A.}\ \bibnamefont
  {Jalabert}}, \ and\ \bibinfo {author} {\bibfnamefont {D.~A.}\ \bibnamefont
  {Wisniacki}},\ }\href {\doibase 10.1103/PhysRevE.100.042201} {\bibfield
  {journal} {\bibinfo  {journal} {Phys. Rev. E}\ }\textbf {\bibinfo {volume}
  {100}},\ \bibinfo {pages} {042201} (\bibinfo {year} {2019})}\BibitemShut
  {NoStop}%
\bibitem [{\citenamefont {Torres-Herrera}\ \emph {et~al.}(2018)\citenamefont
  {Torres-Herrera}, \citenamefont {Garc\'{\i}a-Garc\'{\i}a},\ and\
  \citenamefont {Santos}}]{THerrera2018}%
  \BibitemOpen
  \bibfield  {author} {\bibinfo {author} {\bibfnamefont {E.~J.}\ \bibnamefont
  {Torres-Herrera}}, \bibinfo {author} {\bibfnamefont {A.~M.}\ \bibnamefont
  {Garc\'{\i}a-Garc\'{\i}a}}, \ and\ \bibinfo {author} {\bibfnamefont {L.~F.}\
  \bibnamefont {Santos}},\ }\href {\doibase 10.1103/PhysRevB.97.060303}
  {\bibfield  {journal} {\bibinfo  {journal} {Phys. Rev. B}\ }\textbf {\bibinfo
  {volume} {97}},\ \bibinfo {pages} {060303} (\bibinfo {year}
  {2018})}\BibitemShut {NoStop}%
\bibitem [{\citenamefont {Ch\'avez-Carlos}\ \emph {et~al.}(2019)\citenamefont
  {Ch\'avez-Carlos}, \citenamefont {L\'opez-del Carpio}, \citenamefont
  {Bastarrachea-Magnani}, \citenamefont {Str\'ansk\'y}, \citenamefont
  {Lerma-Hern\'andez}, \citenamefont {Santos},\ and\ \citenamefont
  {Hirsch}}]{CCarlos2019}%
  \BibitemOpen
  \bibfield  {author} {\bibinfo {author} {\bibfnamefont {J.}~\bibnamefont
  {Ch\'avez-Carlos}}, \bibinfo {author} {\bibfnamefont {B.}~\bibnamefont
  {L\'opez-del Carpio}}, \bibinfo {author} {\bibfnamefont {M.~A.}\ \bibnamefont
  {Bastarrachea-Magnani}}, \bibinfo {author} {\bibfnamefont {P.}~\bibnamefont
  {Str\'ansk\'y}}, \bibinfo {author} {\bibfnamefont {S.}~\bibnamefont
  {Lerma-Hern\'andez}}, \bibinfo {author} {\bibfnamefont {L.~F.}\ \bibnamefont
  {Santos}}, \ and\ \bibinfo {author} {\bibfnamefont {J.~G.}\ \bibnamefont
  {Hirsch}},\ }\href {\doibase 10.1103/PhysRevLett.122.024101} {\bibfield
  {journal} {\bibinfo  {journal} {Phys. Rev. Lett.}\ }\textbf {\bibinfo
  {volume} {122}},\ \bibinfo {pages} {024101} (\bibinfo {year}
  {2019})}\BibitemShut {NoStop}%
\bibitem [{\citenamefont {Pilatowsky-Cameo}\ \emph {et~al.}(2019)\citenamefont
  {Pilatowsky-Cameo}, \citenamefont {Chávez-Carlos}, \citenamefont
  {Bastarrachea-Magnani}, \citenamefont {Stránský}, \citenamefont
  {Lerma-Hernández}, \citenamefont {Santos},\ and\ \citenamefont
  {Hirsch}}]{Cameo2019}%
  \BibitemOpen
  \bibfield  {author} {\bibinfo {author} {\bibfnamefont {S.}~\bibnamefont
  {Pilatowsky-Cameo}}, \bibinfo {author} {\bibfnamefont {J.}~\bibnamefont
  {Chávez-Carlos}}, \bibinfo {author} {\bibfnamefont {M.~A.}\ \bibnamefont
  {Bastarrachea-Magnani}}, \bibinfo {author} {\bibfnamefont {P.}~\bibnamefont
  {Stránský}}, \bibinfo {author} {\bibfnamefont {S.}~\bibnamefont
  {Lerma-Hernández}}, \bibinfo {author} {\bibfnamefont {L.~F.}\ \bibnamefont
  {Santos}}, \ and\ \bibinfo {author} {\bibfnamefont {J.~G.}\ \bibnamefont
  {Hirsch}},\ }\href@noop {} {\bibfield  {journal} {\bibinfo  {journal} {arXiv
  e-prints}\ } (\bibinfo {year} {2019})},\ \Eprint
  {http://arxiv.org/abs/1909.02578} {arXiv:1909.02578 [cond-mat.stat-mech]}
  \BibitemShut {NoStop}%
\bibitem [{\citenamefont {Lashkari}\ \emph {et~al.}(2013)\citenamefont
  {Lashkari}, \citenamefont {Stanford}, \citenamefont {Hastings}, \citenamefont
  {Osborne},\ and\ \citenamefont {Hayden}}]{Lashkari2013}%
  \BibitemOpen
  \bibfield  {author} {\bibinfo {author} {\bibfnamefont {N.}~\bibnamefont
  {Lashkari}}, \bibinfo {author} {\bibfnamefont {D.}~\bibnamefont {Stanford}},
  \bibinfo {author} {\bibfnamefont {M.}~\bibnamefont {Hastings}}, \bibinfo
  {author} {\bibfnamefont {T.}~\bibnamefont {Osborne}}, \ and\ \bibinfo
  {author} {\bibfnamefont {P.}~\bibnamefont {Hayden}},\ }\href {\doibase
  10.1007/JHEP04(2013)022} {\bibfield  {journal} {\bibinfo  {journal} {J. High
  Energy Phys.}\ }\textbf {\bibinfo {volume} {2013}},\ \bibinfo {pages} {22}
  (\bibinfo {year} {2013})}\BibitemShut {NoStop}%
\bibitem [{\citenamefont {Shenker}\ and\ \citenamefont
  {Stanford}(2014)}]{Shenker2014}%
  \BibitemOpen
  \bibfield  {author} {\bibinfo {author} {\bibfnamefont {S.~H.}\ \bibnamefont
  {Shenker}}\ and\ \bibinfo {author} {\bibfnamefont {D.}~\bibnamefont
  {Stanford}},\ }\href {\doibase 10.1007/JHEP03(2014)067} {\bibfield  {journal}
  {\bibinfo  {journal} {J. High Energy Phys.}\ }\textbf {\bibinfo {volume}
  {2014}},\ \bibinfo {pages} {67} (\bibinfo {year} {2014})}\BibitemShut
  {NoStop}%
\bibitem [{\citenamefont {Maldacena}\ \emph {et~al.}(2017)\citenamefont
  {Maldacena}, \citenamefont {Stanford},\ and\ \citenamefont
  {Yang}}]{Maldacena2017}%
  \BibitemOpen
  \bibfield  {author} {\bibinfo {author} {\bibfnamefont {J.}~\bibnamefont
  {Maldacena}}, \bibinfo {author} {\bibfnamefont {D.}~\bibnamefont {Stanford}},
  \ and\ \bibinfo {author} {\bibfnamefont {Z.}~\bibnamefont {Yang}},\ }\href
  {\doibase 10.1002/prop.201700034} {\bibfield  {journal} {\bibinfo  {journal}
  {Fortschr. Phys.}\ }\textbf {\bibinfo {volume} {65}},\ \bibinfo {pages}
  {1700034} (\bibinfo {year} {2017})}\BibitemShut {NoStop}%
\bibitem [{\citenamefont {Polchinski}\ and\ \citenamefont
  {Rosenhaus}(2016)}]{Polchinski2016}%
  \BibitemOpen
  \bibfield  {author} {\bibinfo {author} {\bibfnamefont {J.}~\bibnamefont
  {Polchinski}}\ and\ \bibinfo {author} {\bibfnamefont {V.}~\bibnamefont
  {Rosenhaus}},\ }\href {\doibase 10.1007/JHEP04(2016)001} {\bibfield
  {journal} {\bibinfo  {journal} {J. High Energy Phys.}\ }\textbf {\bibinfo
  {volume} {2016}},\ \bibinfo {pages} {1} (\bibinfo {year} {2016})}\BibitemShut
  {NoStop}%
\bibitem [{\citenamefont {Maldacena}\ and\ \citenamefont
  {Stanford}(2016)}]{Maldacena2016b}%
  \BibitemOpen
  \bibfield  {author} {\bibinfo {author} {\bibfnamefont {J.}~\bibnamefont
  {Maldacena}}\ and\ \bibinfo {author} {\bibfnamefont {D.}~\bibnamefont
  {Stanford}},\ }\href {\doibase 10.1103/PhysRevD.94.106002} {\bibfield
  {journal} {\bibinfo  {journal} {Phys. Rev. D}\ }\textbf {\bibinfo {volume}
  {94}},\ \bibinfo {pages} {106002} (\bibinfo {year} {2016})}\BibitemShut
  {NoStop}%
\bibitem [{\citenamefont {Swingle}\ and\ \citenamefont
  {Chowdhury}(2017)}]{Swingle2017}%
  \BibitemOpen
  \bibfield  {author} {\bibinfo {author} {\bibfnamefont {B.}~\bibnamefont
  {Swingle}}\ and\ \bibinfo {author} {\bibfnamefont {D.}~\bibnamefont
  {Chowdhury}},\ }\href {\doibase 10.1103/PhysRevB.95.060201} {\bibfield
  {journal} {\bibinfo  {journal} {Phys. Rev. B}\ }\textbf {\bibinfo {volume}
  {95}},\ \bibinfo {pages} {060201} (\bibinfo {year} {2017})}\BibitemShut
  {NoStop}%
\bibitem [{\citenamefont {Patel}\ and\ \citenamefont
  {Sachdev}(2017)}]{Patel2017}%
  \BibitemOpen
  \bibfield  {author} {\bibinfo {author} {\bibfnamefont {A.~A.}\ \bibnamefont
  {Patel}}\ and\ \bibinfo {author} {\bibfnamefont {S.}~\bibnamefont
  {Sachdev}},\ }\href {\doibase 10.1073/pnas.1618185114} {\bibfield  {journal}
  {\bibinfo  {journal} {Proc. Natl. Acad. Sci.}\ }\textbf {\bibinfo {volume}
  {114}},\ \bibinfo {pages} {1844} (\bibinfo {year} {2017})}\BibitemShut
  {NoStop}%
\bibitem [{\citenamefont {Patel}\ \emph {et~al.}(2017)\citenamefont {Patel},
  \citenamefont {Chowdhury}, \citenamefont {Sachdev},\ and\ \citenamefont
  {Swingle}}]{Patel2017b}%
  \BibitemOpen
  \bibfield  {author} {\bibinfo {author} {\bibfnamefont {A.~A.}\ \bibnamefont
  {Patel}}, \bibinfo {author} {\bibfnamefont {D.}~\bibnamefont {Chowdhury}},
  \bibinfo {author} {\bibfnamefont {S.}~\bibnamefont {Sachdev}}, \ and\
  \bibinfo {author} {\bibfnamefont {B.}~\bibnamefont {Swingle}},\ }\href
  {\doibase 10.1103/PhysRevX.7.031047} {\bibfield  {journal} {\bibinfo
  {journal} {Phys. Rev. X}\ }\textbf {\bibinfo {volume} {7}},\ \bibinfo {pages}
  {031047} (\bibinfo {year} {2017})}\BibitemShut {NoStop}%
\bibitem [{\citenamefont {D\'ora}\ and\ \citenamefont
  {Moessner}(2017)}]{Dora2017}%
  \BibitemOpen
  \bibfield  {author} {\bibinfo {author} {\bibfnamefont {B.}~\bibnamefont
  {D\'ora}}\ and\ \bibinfo {author} {\bibfnamefont {R.}~\bibnamefont
  {Moessner}},\ }\href {\doibase 10.1103/PhysRevLett.119.026802} {\bibfield
  {journal} {\bibinfo  {journal} {Phys. Rev. Lett.}\ }\textbf {\bibinfo
  {volume} {119}},\ \bibinfo {pages} {026802} (\bibinfo {year}
  {2017})}\BibitemShut {NoStop}%
\bibitem [{\citenamefont {Shen}\ \emph {et~al.}(2017)\citenamefont {Shen},
  \citenamefont {Zhang}, \citenamefont {Fan},\ and\ \citenamefont
  {Zhai}}]{Shen2017}%
  \BibitemOpen
  \bibfield  {author} {\bibinfo {author} {\bibfnamefont {H.}~\bibnamefont
  {Shen}}, \bibinfo {author} {\bibfnamefont {P.}~\bibnamefont {Zhang}},
  \bibinfo {author} {\bibfnamefont {R.}~\bibnamefont {Fan}}, \ and\ \bibinfo
  {author} {\bibfnamefont {H.}~\bibnamefont {Zhai}},\ }\href {\doibase
  10.1103/PhysRevB.96.054503} {\bibfield  {journal} {\bibinfo  {journal} {Phys.
  Rev. B}\ }\textbf {\bibinfo {volume} {96}},\ \bibinfo {pages} {054503}
  (\bibinfo {year} {2017})}\BibitemShut {NoStop}%
\bibitem [{\citenamefont {{Sun}}\ \emph {et~al.}(2018)\citenamefont {{Sun}},
  \citenamefont {{Cai}}, \citenamefont {{Tang}}, \citenamefont {{Hu}},\ and\
  \citenamefont {{Fan}}}]{Sunz2018}%
  \BibitemOpen
  \bibfield  {author} {\bibinfo {author} {\bibfnamefont {Z.-H.}\ \bibnamefont
  {{Sun}}}, \bibinfo {author} {\bibfnamefont {J.-Q.}\ \bibnamefont {{Cai}}},
  \bibinfo {author} {\bibfnamefont {Q.-C.}\ \bibnamefont {{Tang}}}, \bibinfo
  {author} {\bibfnamefont {Y.}~\bibnamefont {{Hu}}}, \ and\ \bibinfo {author}
  {\bibfnamefont {H.}~\bibnamefont {{Fan}}},\ }\href@noop {} {\bibfield
  {journal} {\bibinfo  {journal} {arXiv e-prints}\ } (\bibinfo {year}
  {2018})},\ \Eprint {http://arxiv.org/abs/1811.11191} {arXiv:1811.11191
  [quant-ph]} \BibitemShut {NoStop}%
\bibitem [{\citenamefont {Heyl}\ \emph {et~al.}(2018)\citenamefont {Heyl},
  \citenamefont {Pollmann},\ and\ \citenamefont {D\'ora}}]{Heyl2018}%
  \BibitemOpen
  \bibfield  {author} {\bibinfo {author} {\bibfnamefont {M.}~\bibnamefont
  {Heyl}}, \bibinfo {author} {\bibfnamefont {F.}~\bibnamefont {Pollmann}}, \
  and\ \bibinfo {author} {\bibfnamefont {B.}~\bibnamefont {D\'ora}},\ }\href
  {\doibase 10.1103/PhysRevLett.121.016801} {\bibfield  {journal} {\bibinfo
  {journal} {Phys. Rev. Lett.}\ }\textbf {\bibinfo {volume} {121}},\ \bibinfo
  {pages} {016801} (\bibinfo {year} {2018})}\BibitemShut {NoStop}%
\bibitem [{\citenamefont {Huang}\ \emph {et~al.}(2017)\citenamefont {Huang},
  \citenamefont {Zhang},\ and\ \citenamefont {Chen}}]{Huang2017}%
  \BibitemOpen
  \bibfield  {author} {\bibinfo {author} {\bibfnamefont {Y.}~\bibnamefont
  {Huang}}, \bibinfo {author} {\bibfnamefont {Y.-L.}\ \bibnamefont {Zhang}}, \
  and\ \bibinfo {author} {\bibfnamefont {X.}~\bibnamefont {Chen}},\ }\href
  {\doibase 10.1002/andp.201600318} {\bibfield  {journal} {\bibinfo  {journal}
  {Ann. Phys.-Berlin}\ }\textbf {\bibinfo {volume} {529}},\ \bibinfo {pages}
  {1600318} (\bibinfo {year} {2017})}\BibitemShut {NoStop}%
\bibitem [{\citenamefont {Fan}\ \emph {et~al.}(2017)\citenamefont {Fan},
  \citenamefont {Zhang}, \citenamefont {Shen},\ and\ \citenamefont
  {Zhai}}]{Fan2017}%
  \BibitemOpen
  \bibfield  {author} {\bibinfo {author} {\bibfnamefont {R.}~\bibnamefont
  {Fan}}, \bibinfo {author} {\bibfnamefont {P.}~\bibnamefont {Zhang}}, \bibinfo
  {author} {\bibfnamefont {H.}~\bibnamefont {Shen}}, \ and\ \bibinfo {author}
  {\bibfnamefont {H.}~\bibnamefont {Zhai}},\ }\href {\doibase
  https://doi.org/10.1016/j.scib.2017.04.011} {\bibfield  {journal} {\bibinfo
  {journal} {Sci. Bull.}\ }\textbf {\bibinfo {volume} {62}},\ \bibinfo {pages}
  {707 } (\bibinfo {year} {2017})}\BibitemShut {NoStop}%
\bibitem [{\citenamefont {Sahu}\ \emph {et~al.}(2019)\citenamefont {Sahu},
  \citenamefont {Xu},\ and\ \citenamefont {Swingle}}]{Sahu2018}%
  \BibitemOpen
  \bibfield  {author} {\bibinfo {author} {\bibfnamefont {S.}~\bibnamefont
  {Sahu}}, \bibinfo {author} {\bibfnamefont {S.}~\bibnamefont {Xu}}, \ and\
  \bibinfo {author} {\bibfnamefont {B.}~\bibnamefont {Swingle}},\ }\href
  {\doibase 10.1103/PhysRevLett.123.165902} {\bibfield  {journal} {\bibinfo
  {journal} {Phys. Rev. Lett.}\ }\textbf {\bibinfo {volume} {123}},\ \bibinfo
  {pages} {165902} (\bibinfo {year} {2019})}\BibitemShut {NoStop}%
\bibitem [{\citenamefont {Campisi}\ and\ \citenamefont
  {Goold}(2017)}]{Campisi2017}%
  \BibitemOpen
  \bibfield  {author} {\bibinfo {author} {\bibfnamefont {M.}~\bibnamefont
  {Campisi}}\ and\ \bibinfo {author} {\bibfnamefont {J.}~\bibnamefont
  {Goold}},\ }\href {\doibase 10.1103/PhysRevE.95.062127} {\bibfield  {journal}
  {\bibinfo  {journal} {Phys. Rev. E}\ }\textbf {\bibinfo {volume} {95}},\
  \bibinfo {pages} {062127} (\bibinfo {year} {2017})}\BibitemShut {NoStop}%
\bibitem [{\citenamefont {Chenu}\ \emph {et~al.}(2018)\citenamefont {Chenu},
  \citenamefont {Egusquiza}, \citenamefont {Molina-Vilaplana},\ and\
  \citenamefont {del Campo}}]{Chenu2018}%
  \BibitemOpen
  \bibfield  {author} {\bibinfo {author} {\bibfnamefont {A.}~\bibnamefont
  {Chenu}}, \bibinfo {author} {\bibfnamefont {I.~L.}\ \bibnamefont
  {Egusquiza}}, \bibinfo {author} {\bibfnamefont {J.}~\bibnamefont
  {Molina-Vilaplana}}, \ and\ \bibinfo {author} {\bibfnamefont
  {A.}~\bibnamefont {del Campo}},\ }\href {\doibase
  https://arxiv.org/ct?url=https\%3A\%2F\%2Fdx.doi.org\%2F10.1038\%2Fs41598-018-30982-w&v=55b43499}
  {\bibfield  {journal} {\bibinfo  {journal} {Sci. Rep.}\ }\textbf {\bibinfo
  {volume} {8}},\ \bibinfo {pages} {12634} (\bibinfo {year}
  {2018})}\BibitemShut {NoStop}%
\bibitem [{\citenamefont {Chenu}\ \emph {et~al.}(2019)\citenamefont {Chenu},
  \citenamefont {Molina-Vilaplana},\ and\ \citenamefont {del
  Campo}}]{Chenu2019}%
  \BibitemOpen
  \bibfield  {author} {\bibinfo {author} {\bibfnamefont {A.}~\bibnamefont
  {Chenu}}, \bibinfo {author} {\bibfnamefont {J.}~\bibnamefont
  {Molina-Vilaplana}}, \ and\ \bibinfo {author} {\bibfnamefont
  {A.}~\bibnamefont {del Campo}},\ }\href {\doibase
  https://doi.org/10.22331/q-2019-03-04-127} {\bibfield  {journal} {\bibinfo
  {journal} {Quantum}\ }\textbf {\bibinfo {volume} {3}},\ \bibinfo {pages}
  {127} (\bibinfo {year} {2019})}\BibitemShut {NoStop}%
\bibitem [{\citenamefont {Buijsman}\ \emph {et~al.}(2017)\citenamefont
  {Buijsman}, \citenamefont {Gritsev},\ and\ \citenamefont
  {Sprik}}]{Buijsman2017}%
  \BibitemOpen
  \bibfield  {author} {\bibinfo {author} {\bibfnamefont {W.}~\bibnamefont
  {Buijsman}}, \bibinfo {author} {\bibfnamefont {V.}~\bibnamefont {Gritsev}}, \
  and\ \bibinfo {author} {\bibfnamefont {R.}~\bibnamefont {Sprik}},\ }\href
  {\doibase 10.1103/PhysRevLett.118.080601} {\bibfield  {journal} {\bibinfo
  {journal} {Phys. Rev. Lett.}\ }\textbf {\bibinfo {volume} {118}},\ \bibinfo
  {pages} {080601} (\bibinfo {year} {2017})}\BibitemShut {NoStop}%
\bibitem [{\citenamefont {Ray}\ \emph {et~al.}(2018)\citenamefont {Ray},
  \citenamefont {Sinha},\ and\ \citenamefont {Sengupta}}]{Ray2018}%
  \BibitemOpen
  \bibfield  {author} {\bibinfo {author} {\bibfnamefont {S.}~\bibnamefont
  {Ray}}, \bibinfo {author} {\bibfnamefont {S.}~\bibnamefont {Sinha}}, \ and\
  \bibinfo {author} {\bibfnamefont {K.}~\bibnamefont {Sengupta}},\ }\href
  {\doibase 10.1103/PhysRevA.98.053631} {\bibfield  {journal} {\bibinfo
  {journal} {Phys. Rev. A}\ }\textbf {\bibinfo {volume} {98}},\ \bibinfo
  {pages} {053631} (\bibinfo {year} {2018})}\BibitemShut {NoStop}%
\bibitem [{\citenamefont {Li}\ \emph {et~al.}(2017)\citenamefont {Li},
  \citenamefont {Fan}, \citenamefont {Wang}, \citenamefont {Ye}, \citenamefont
  {Zeng}, \citenamefont {Zhai}, \citenamefont {Peng},\ and\ \citenamefont
  {Du}}]{Li2017}%
  \BibitemOpen
  \bibfield  {author} {\bibinfo {author} {\bibfnamefont {J.}~\bibnamefont
  {Li}}, \bibinfo {author} {\bibfnamefont {R.}~\bibnamefont {Fan}}, \bibinfo
  {author} {\bibfnamefont {H.}~\bibnamefont {Wang}}, \bibinfo {author}
  {\bibfnamefont {B.}~\bibnamefont {Ye}}, \bibinfo {author} {\bibfnamefont
  {B.}~\bibnamefont {Zeng}}, \bibinfo {author} {\bibfnamefont {H.}~\bibnamefont
  {Zhai}}, \bibinfo {author} {\bibfnamefont {X.}~\bibnamefont {Peng}}, \ and\
  \bibinfo {author} {\bibfnamefont {J.}~\bibnamefont {Du}},\ }\href {\doibase
  10.1103/PhysRevX.7.031011} {\bibfield  {journal} {\bibinfo  {journal} {Phys.
  Rev. X}\ }\textbf {\bibinfo {volume} {7}},\ \bibinfo {pages} {031011}
  (\bibinfo {year} {2017})}\BibitemShut {NoStop}%
\bibitem [{\citenamefont {Wei}\ \emph {et~al.}(2018)\citenamefont {Wei},
  \citenamefont {Ramanathan},\ and\ \citenamefont {Cappellaro}}]{Wei2018}%
  \BibitemOpen
  \bibfield  {author} {\bibinfo {author} {\bibfnamefont {K.~X.}\ \bibnamefont
  {Wei}}, \bibinfo {author} {\bibfnamefont {C.}~\bibnamefont {Ramanathan}}, \
  and\ \bibinfo {author} {\bibfnamefont {P.}~\bibnamefont {Cappellaro}},\
  }\href {\doibase 10.1103/PhysRevLett.120.070501} {\bibfield  {journal}
  {\bibinfo  {journal} {Phys. Rev. Lett.}\ }\textbf {\bibinfo {volume} {120}},\
  \bibinfo {pages} {070501} (\bibinfo {year} {2018})}\BibitemShut {NoStop}%
\bibitem [{\citenamefont {{G{\"a}rttner}}\ \emph {et~al.}(2017)\citenamefont
  {{G{\"a}rttner}}, \citenamefont {{Bohnet}}, \citenamefont {{Safavi-Naini}},
  \citenamefont {{Wall}}, \citenamefont {{Bollinger}},\ and\ \citenamefont
  {{Rey}}}]{Garttner2017}%
  \BibitemOpen
  \bibfield  {author} {\bibinfo {author} {\bibfnamefont {M.}~\bibnamefont
  {{G{\"a}rttner}}}, \bibinfo {author} {\bibfnamefont {J.~G.}\ \bibnamefont
  {{Bohnet}}}, \bibinfo {author} {\bibfnamefont {A.}~\bibnamefont
  {{Safavi-Naini}}}, \bibinfo {author} {\bibfnamefont {M.~L.}\ \bibnamefont
  {{Wall}}}, \bibinfo {author} {\bibfnamefont {J.~J.}\ \bibnamefont
  {{Bollinger}}}, \ and\ \bibinfo {author} {\bibfnamefont {A.~M.}\ \bibnamefont
  {{Rey}}},\ }\href {\doibase 10.1038/nphys4119} {\bibfield  {journal}
  {\bibinfo  {journal} {Nat. Phys.}\ }\textbf {\bibinfo {volume} {13}},\
  \bibinfo {pages} {781} (\bibinfo {year} {2017})}\BibitemShut {NoStop}%
\bibitem [{\citenamefont {Borgonovi}\ \emph {et~al.}(2019)\citenamefont
  {Borgonovi}, \citenamefont {Izrailev},\ and\ \citenamefont
  {Santos}}]{Borgonovi2019}%
  \BibitemOpen
  \bibfield  {author} {\bibinfo {author} {\bibfnamefont {F.}~\bibnamefont
  {Borgonovi}}, \bibinfo {author} {\bibfnamefont {F.~M.}\ \bibnamefont
  {Izrailev}}, \ and\ \bibinfo {author} {\bibfnamefont {L.~F.}\ \bibnamefont
  {Santos}},\ }\href {\doibase 10.1103/PhysRevE.99.052143} {\bibfield
  {journal} {\bibinfo  {journal} {Phys. Rev. E}\ }\textbf {\bibinfo {volume}
  {99}},\ \bibinfo {pages} {052143} (\bibinfo {year} {2019})}\BibitemShut
  {NoStop}%
\bibitem [{\citenamefont {Peres}(1984)}]{Peres1984}%
  \BibitemOpen
  \bibfield  {author} {\bibinfo {author} {\bibfnamefont {A.}~\bibnamefont
  {Peres}},\ }\href {\doibase 10.1103/PhysRevA.30.1610} {\bibfield  {journal}
  {\bibinfo  {journal} {Phys. Rev. A}\ }\textbf {\bibinfo {volume} {30}},\
  \bibinfo {pages} {1610} (\bibinfo {year} {1984})}\BibitemShut {NoStop}%
\bibitem [{\citenamefont {Gu}\ and\ \citenamefont {Qi}(2016)}]{Gu2016}%
  \BibitemOpen
  \bibfield  {author} {\bibinfo {author} {\bibfnamefont {Y.}~\bibnamefont
  {Gu}}\ and\ \bibinfo {author} {\bibfnamefont {X.-L.}\ \bibnamefont {Qi}},\
  }\href {\doibase 10.1007/JHEP08(2016)129} {\bibfield  {journal} {\bibinfo
  {journal} {Journal of High Energy Physics}\ }\textbf {\bibinfo {volume}
  {2016}},\ \bibinfo {pages} {129} (\bibinfo {year} {2016})}\BibitemShut
  {NoStop}%
\bibitem [{\citenamefont {Lipkin}\ \emph {et~al.}(1965)\citenamefont {Lipkin},
  \citenamefont {Meshkov},\ and\ \citenamefont {Glick}}]{Lipkin1965}%
  \BibitemOpen
  \bibfield  {author} {\bibinfo {author} {\bibfnamefont {H.~J.}\ \bibnamefont
  {Lipkin}}, \bibinfo {author} {\bibfnamefont {N.}~\bibnamefont {Meshkov}}, \
  and\ \bibinfo {author} {\bibfnamefont {A.~J.}\ \bibnamefont {Glick}},\
  }\href@noop {} {\bibfield  {journal} {\bibinfo  {journal} {Nucl. Phys.}\
  }\textbf {\bibinfo {volume} {62}},\ \bibinfo {pages} {188} (\bibinfo {year}
  {1965})}\BibitemShut {NoStop}%
\bibitem [{\citenamefont {Dusuel}\ and\ \citenamefont
  {Vidal}(2004)}]{Dusuel2004}%
  \BibitemOpen
  \bibfield  {author} {\bibinfo {author} {\bibfnamefont {S.}~\bibnamefont
  {Dusuel}}\ and\ \bibinfo {author} {\bibfnamefont {J.}~\bibnamefont {Vidal}},\
  }\href@noop {} {\bibfield  {journal} {\bibinfo  {journal} {Phys. Rev. Lett.}\
  }\textbf {\bibinfo {volume} {93}},\ \bibinfo {pages} {237204} (\bibinfo
  {year} {2004})}\BibitemShut {NoStop}%
\bibitem [{\citenamefont {Leyvraz}\ and\ \citenamefont
  {Heiss}(2005)}]{Leyvraz2005}%
  \BibitemOpen
  \bibfield  {author} {\bibinfo {author} {\bibfnamefont {F.}~\bibnamefont
  {Leyvraz}}\ and\ \bibinfo {author} {\bibfnamefont {W.~D.}\ \bibnamefont
  {Heiss}},\ }\href@noop {} {\bibfield  {journal} {\bibinfo  {journal} {Phys.
  Rev. Lett.}\ }\textbf {\bibinfo {volume} {95}},\ \bibinfo {pages} {050402}
  (\bibinfo {year} {2005})}\BibitemShut {NoStop}%
\bibitem [{\citenamefont {Casta\~nos}\ \emph {et~al.}(2006)\citenamefont
  {Casta\~nos}, \citenamefont {L\'opez-Pe\~na}, \citenamefont {Hirsch},\ and\
  \citenamefont {L\'opez-Moreno}}]{ocasta1}%
  \BibitemOpen
  \bibfield  {author} {\bibinfo {author} {\bibfnamefont {O.}~\bibnamefont
  {Casta\~nos}}, \bibinfo {author} {\bibfnamefont {R.}~\bibnamefont
  {L\'opez-Pe\~na}}, \bibinfo {author} {\bibfnamefont {J.~G.}\ \bibnamefont
  {Hirsch}}, \ and\ \bibinfo {author} {\bibfnamefont {E.}~\bibnamefont
  {L\'opez-Moreno}},\ }\href {\doibase 10.1103/PhysRevB.74.104118} {\bibfield
  {journal} {\bibinfo  {journal} {Phys. Rev. B}\ }\textbf {\bibinfo {volume}
  {74}},\ \bibinfo {pages} {104118} (\bibinfo {year} {2006})}\BibitemShut
  {NoStop}%
\bibitem [{\citenamefont {Botet}\ and\ \citenamefont
  {Jullien}(1983)}]{Botet1983}%
  \BibitemOpen
  \bibfield  {author} {\bibinfo {author} {\bibfnamefont {R.}~\bibnamefont
  {Botet}}\ and\ \bibinfo {author} {\bibfnamefont {R.}~\bibnamefont
  {Jullien}},\ }\href {\doibase 10.1103/PhysRevB.28.3955} {\bibfield  {journal}
  {\bibinfo  {journal} {Phys. Rev. B}\ }\textbf {\bibinfo {volume} {28}},\
  \bibinfo {pages} {3955} (\bibinfo {year} {1983})}\BibitemShut {NoStop}%
\bibitem [{\citenamefont {P\'erez-Fern\'andez}\ and\ \citenamefont
  {Rela\~no}(2017)}]{PPerez2017}%
  \BibitemOpen
  \bibfield  {author} {\bibinfo {author} {\bibfnamefont {P.}~\bibnamefont
  {P\'erez-Fern\'andez}}\ and\ \bibinfo {author} {\bibfnamefont
  {A.}~\bibnamefont {Rela\~no}},\ }\href {\doibase 10.1103/PhysRevE.96.012121}
  {\bibfield  {journal} {\bibinfo  {journal} {Phys. Rev. E}\ }\textbf {\bibinfo
  {volume} {96}},\ \bibinfo {pages} {012121} (\bibinfo {year}
  {2017})}\BibitemShut {NoStop}%
\bibitem [{\citenamefont {Da\ifmmode~\breve{g}\else \u{g}\fi{}}\ \emph
  {et~al.}(2019)\citenamefont {Da\ifmmode~\breve{g}\else \u{g}\fi{}},
  \citenamefont {Sun},\ and\ \citenamefont {Duan}}]{Dag2019}%
  \BibitemOpen
  \bibfield  {author} {\bibinfo {author} {\bibfnamefont {C.~B.}\ \bibnamefont
  {Da\ifmmode~\breve{g}\else \u{g}\fi{}}}, \bibinfo {author} {\bibfnamefont
  {K.}~\bibnamefont {Sun}}, \ and\ \bibinfo {author} {\bibfnamefont {L.-M.}\
  \bibnamefont {Duan}},\ }\href {\doibase 10.1103/PhysRevLett.123.140602}
  {\bibfield  {journal} {\bibinfo  {journal} {Phys. Rev. Lett.}\ }\textbf
  {\bibinfo {volume} {123}},\ \bibinfo {pages} {140602} (\bibinfo {year}
  {2019})}\BibitemShut {NoStop}%
\end{thebibliography}%

\end{document}